\documentclass[12pt]{qbio}

\usepackage{graphicx}
\usepackage{dcolumn}
\usepackage{bm}
\usepackage{hyperref}
\usepackage{latexsym}
\usepackage{amssymb}
\usepackage{lineno}
\usepackage{times}

\usepackage{natbib}

\begin{document}

\runhead{Patterns in food web structure}
\pagestyle{headings}

\begin{frontmatter}

\title{Quantitative patterns in the structure of model and empirical food webs}

\author{J. Camacho}$^{1,2}$\thanks{{\bf Email}: Juan.Camacho@uab.es,
rguimera@northwestern.edu, d-stouffer@northwestern.edu, and\\
amaral@northwestern.edu},
\author{R. Guimer\`a}$^{1}$,
\author{D. B. Stouffer}$^{1}$, and
\author{L. A. N. Amaral}$^{1}$
\address{
$^1$ Department of Chemical and Biological Engineering, Northwestern
University, \\Evanston, IL 60208, USA \\
$^2$ Departament de F\'{\i}sica (F\'{\i}sica Estad\'{\i}stica),
Universitat Aut\`onoma de Barcelona, \\E-08193 Bellaterra, Catalonia,
Spain
}

\end{frontmatter}

\thispagestyle{empty}

\clearpage

\centerline{\Large\bf Abstract}
\vspace{-0.3cm}
We analyze the properties of model food webs and of fifteen community
food webs from a variety of environments ---including freshwater,
marine-freshwater interfaces and terrestrial environments. 
We first perform a theoretical analysis of a recently proposed model
for food webs---the niche model of Williams and Martinez
(\citeyear{martinez00}). We derive analytical expressions for the
distributions of species' number of prey, number of predators, and
total number of trophic links and find that they follow universal
functional forms. We also derive expressions for a number of other
biologically relevant parameters which depend on these distributions.
These include the fraction of top, intermediate, basal, and cannibal
species, the standard deviations of generality and vulnerability, the
correlation coefficient between species' number of prey and number of
predators, and assortativity. We show that our findings are robust
under rather general conditions; a result which could not have been
demonstrated without treating the problem analytically.
We then use our analytical predictions as a guide to the analysis of
fifteen of the most complete empirical food webs available.  We
uncover quantitative unifying patterns that describe the properties of
the model food webs and most of the trophic webs considered.  Our
results support a strong new hypothesis that the empirical
distributions of number of prey and number of predators follow {\it
universal functional forms that, without free parameters, match our
analytical predictions}. Further, we find that the empirically
observed correlation coefficient, assortativity, and fraction of
cannibal species are consistent with our analytical expressions and
simulations of the niche model.  Finally, we show that two quantities
typically used to characterize complex networks, the average distance
between nodes and the average clustering coefficient of the nodes,
show a high degree of regularity for both the empirical data and
simulations of the niche model.
Our findings suggest that statistical physics concepts such as scaling
and universality may be useful in the description of natural
ecosystems.

\vfill


\clearpage

\section{Introduction}

In ecosystems, species are connected through intricate trophic
relationships which define complex food webs
\citep{briand84,cohen90,martinez00}. Understanding the structure and
mechanisms underlying the formation of these complex webs is of great
importance in ecology \citep{rejmanek79,briand87,cohen90,polis91}. In
particular, food web structure provides insights into the behavior of
ecosystems under perturbations such as the introduction of new species
or the extinction of existing ones. It is thought that the nonlinear
response of the elements composing the network may lead to possibly
catastrophic outcomes for even small perturbations
\citep{berlow99,chapin00,mckann00}. Thus, an understanding of the
topology and assembly mechanisms of food webs may be of great
practical interest with regard to crop selection, management of
fishing stocks, preservation of threatened ecosystems, and maintenance
of biodiversity \citep{pimm95,stone99,covington00,hutchings00}.

Due to the importance of food webs, much effort has been put into
collecting empirical data and uncovering unifying patterns which
describe their structure
\citep{rejmanek79,briand84,briand87,lawton88,cohen90,hall91,pimm91,polis91,martinez93}.
However, in the last decade the construction of larger and more
complete food webs has clearly demonstrated that the previously
reported unifying patterns do not hold for the new webs
\citep{hall91,hall93,polis91}. Indeed, the complexity of the new webs
has rendered it quite challenging to obtain quantitative patterns with
which to ``substitute'' the old ones \citep{dunne02}.

The purpose of this work is twofold. First, to perform a theoretical
analysis of a recently proposed model for food webs---the niche model
of Williams and Martinez (\citeyear{martinez00}). Second, to use the
predictions of the model as a guide to a systematic statistical
analysis of community food webs from a variety of environments.

Remarkably, we uncover quantitative unifying patterns that describe
the properties of most of the diverse trophic webs considered and
capture the random and non-random aspects of their
structure. Specifically, we find that several quantities---such as the
distributions of number of prey, number of predators, and number of
trophic links---characterizing these diverse food webs obey robust
functional forms that, as predicted by our analytical results for the
niche model, depend on a single parameter---the linkage density of the
food web.

The organization of the paper is as follows. In Section 2 we study the
niche model proposed by Williams and Martinez (\citeyear{martinez00})
analytically and numerically. In Section 3 we analyze the empirical
food webs and show the existence of robust quantitative
patterns. Finally, in Section 4 we present some concluding remarks.

\section{Analytical solution of the niche model}

Recently, Williams and Martinez (\citeyear{martinez00}) have proposed
a model for food web structure---the niche model---that with just a
couple of ingredients appears to successfully predict key structural
properties of the most comprehensive food webs in the
literature. Numerical simulations of the niche model predicted values
for many quantities typically used to characterize empirical food webs
that are consistent with measured values for seven empirical webs.


In this Section, we systematically investigate the niche model from a
{\it theoretical} perspective. We obtain analytical expressions for a
number of quantities characterizing the structure of food webs in the
limit of sparse food webs, i.e., webs with $L \ll S^2$, where $L$ is
the number of trophic interactions between species and $S$ is the
number of species in the web. This limit is the one of interest in
ecology because (i) for most food webs reported in the literature the
directed connectance, defined as $C \equiv L/S^2$, take values much
smaller than one, and (ii) it corresponds to the limit of large web
sizes $S$ \citep{briand84,briand87,sole02}, which is the interesting
limit if one surmises that geographically separate ecosystems are in
fact connected.

We first calculate the probability distributions of number of prey and
of number of predators and find that for $C \ll 1$ they depend only on
one parameter of the model---the linkage density $z \equiv L / S$ ,
i.e. the average number of prey or predators. These distributions give
valuable information about the structure of the network
\citep{albert00} and enable us to calculate other interesting
quantities such as the fraction of ``top,'' ``intermediate,'' and
``basal'' species, and the standard deviation of the ``vulnerability''
and ``generality'' of the species in the food web
\citep{martinez00}. We also calculate the correlation coefficient
between number of prey and number of predators, the fraction of
cannibals present, and two additional properties of interest in the
characterization of complex networks: the average ``distance'' between
species and the local redundancy of the connection between species
\citep{watts98}.

\subsection{The niche model}
Consider an ecosystem with $S$ species and $L$ trophic interactions
between these species. These species and interactions define a
network with $S$ nodes and $L$ directed links. In the niche model, one
first randomly assigns $S$ species to ``trophic niches'' with niche
values $n_i$ mapped uniformly onto the interval [0,1]. For
convenience, we will assume that the species are ordered according to
their niche number, i.e., $n_1 < n_2 < ... < n_S$.

A species $i$ is characterized by its niche parameter $n_i$ and by its
list of prey. Prey are chosen for all species according to the
following rule: A species $i$ preys on all species $j$ with niche
parameter $n_j$ inside a segment of length $a_i$ centered in a
position chosen randomly inside the interval $[a_i/2,n_i]$, with $a_i
= x n_i$ and $0 \le x \le 1$ a random variable with probability
density function
\begin{equation}
\label{px}
p_{x}(x) = b \left(1-x\right) ^{\left( b-1\right)}\,.
\end{equation}
Williams and Martinez (\citeyear{martinez00}) appeared to have chosen
this functional form for convenience, but, we will show later that the
predictions of the model are mostly robust to changes in the specific
form of $p_{x}$.

The values of the parameters $b$ and $S$ determine the linkage density
$z= L/S$ of the food web, and the directed connectance
$C=L/S^2$. Since the species are uniformly distributed along the
segment $[0,1]$, one can express the average number of prey per
species as $S\overline{a}$, where the bar indicates an average over
all species in the web. It then follows that the linkage density is
\begin{equation}
\label{zvalue}
z=S\overline{a}\,
\end{equation}
and the connectance is 
\begin{equation}
\label{Cvalue}
C = \overline{a}\,.
\end{equation}
One can also obtain these expressions in terms of $b$ by noting that
$n$ and $x$ are independent variables, namely
\begin{equation}
\label{ravg}
\overline{a}=\overline{n}\, \overline{x}={1 \over{2\left(1+b\right)}}\,,
\end{equation}
where $\overline{n}=1/2$ and $\overline{x}=1/\left(1+b\right)$. 

In the niche model, isolated species---that is, species with no prey
or predators---are eliminated, and species with the same list of prey
and predators---that is, trophically-identical species---are
``merged.'' For $C=\overline{a} \ll 1$, the probability that two
species are trophically identical is very small. This suggests that
taxonomic and trophic classifications of species lead to similar
results in this limit \citep{martinez00}.

In the following subsections we derive analytical expressions that
enable us to predict the properties of the webs generated by model.

\subsection{Distribution of number of prey}

\begin{figure}[t]
\centerline{\includegraphics*[width=0.5\columnwidth]{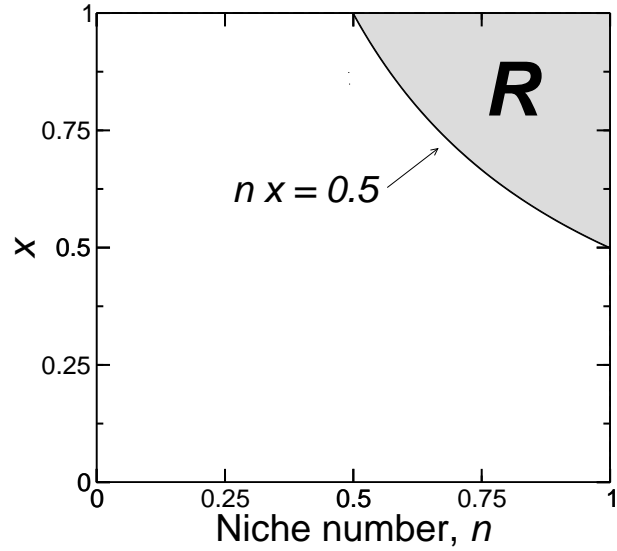}}
\caption{Calculation of the distribution of segment lengths $a$. Each
point inside the $n-x$ diagram yields a value $a = n x$. The hyperbole
$n x=$ const divides the $n-x$ plane in two regions, which we color
grey and white. The region $R$, colored in grey, contains all the
points for which $a>0.5$. Therefore, the probability of drawing a pair
$(n,x)$ belonging to region $R$ can be calculated using
Eq.~(\protect\ref{cump}).}
\label{fig0}
\end{figure}

For large $S$, the number of prey of species $i$ is $k_i = S a_i$, so
that the probability of having $k$ prey $p_{\rm prey}$ is given
directly by the distribution of $a$. Specifically,
\begin{equation}
\label{preyfromr}
p_{\rm prey}(k) = p(a) / S\,.
\end{equation}
In order to evaluate $p(a)$, let us note that $a$ is the product of
two independent stochastic variables, $n$ and $x$, both ranging
between $0$ and $1$. As illustrated in Fig.~\ref{fig0}, it then
follows that the cumulative probability $P(a'>a) = \int
_{a}^{1}da'p(a')$ is the probability that a pair of random values
$(n,x)$ fall in the region $R$ of the $n-x$ diagram bounded by the
lines $x = 1$, $n = 1$, and the hyperbole $a = nx$,
\begin{eqnarray}
\label{cump}
 P(a'>a) &=& \int _{a}^{1}da'p(a') \nonumber\\
 &=& \int_R \, dx\, dn\, p_{n}(n)\, p_{x}(x)\nonumber\\
 &=&\int_a^1\, dx\, \int_{a/x}^1\, dn\, p_{n}(n)\, p_{x}(x)\,,
\end{eqnarray}
where $p_{n}(n) = 1$ is the probability density function for $n$. The
integration of Eq.~(\ref{cump}) gives rise to a closed form involving
hypergeometric functions \citep{gradstheyn00}.

However, we can obtain a simpler analytical solution for $p(a)$ than
that given by the hypergeometric functions as follows. In the limit $C
\ll 1$, which implies $b \gg 1$, $p_{x}(x)$ is negligible except when
$x \ll 1$. We can then approximate $p_x$ as
\begin{equation}
\label{px_xsmall}
p_{x}(x) = b \left(1-x\right) ^{\left( b-1\right)} \simeq b e^{-b x}\,.
\end{equation}
in the entire $x$-range and expect the results to remain unchanged in
the limit $C \ll 1$. Under this approximation, the solution of
Eq.~(\ref{cump}) yields
\begin{equation}
\label{presas}
p(a)=bE_{1}(ba) \,,
\end{equation}
where $E_{1}(x) = \int ^{\infty }_{x} dt~t^{-1} \exp(-t)$ is the
exponential-integral function \citep{gradstheyn00}. The probability
distribution $p_{\rm prey}(k)$ is obtained from Eq.~(\ref{presas})
making the substitutions $a = k/S$ and $b = S/2z$---which are valid in
the limit $C \ll 1$---yielding
\begin{equation}
\label{prey}
p_{\rm prey}(k)=(1/2z)\, E_{1}(k/2z) \,.
\end{equation}

\begin{figure}[t]
\centerline{\includegraphics*[width=\columnwidth]{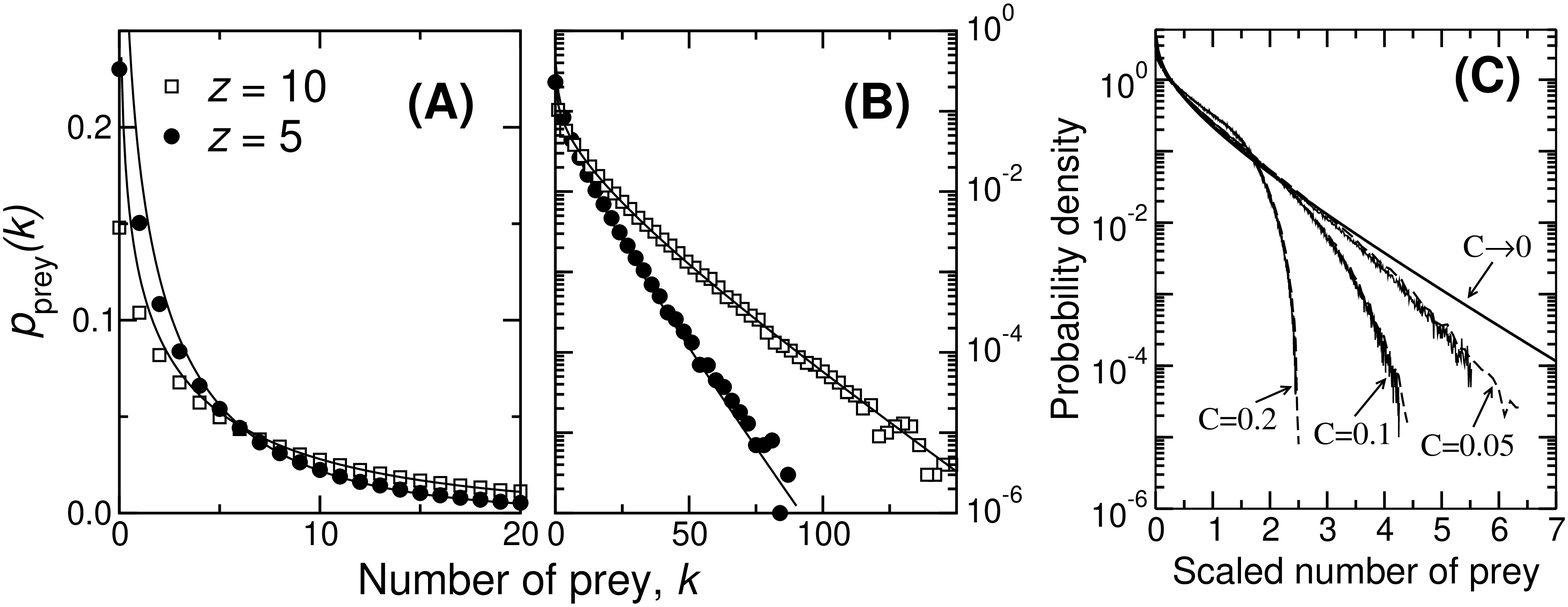}}
\caption{(A) Linear and (B) log-linear plots of the distribution of
 number of prey for 1000 simulations of food webs with $S=1000$. We
 show results for $z=5$ and $10$ and the corresponding theoretical
 predictions. As predicted by Eq.~(\ref{prey}), we find an exponential
 decay of the distributions. (C) Log-linear plot of the probability
 densities for the scaled number of prey $k/2z$ for finite $C$. We
 show results for $C=0.05, 0.1$ and $0.2,$ for $50000$ webs generated
 using the niche model. Dashed lines (solid lines) correspond to
 simulations with $S=100$ ($S=1000$) for the appropriate values of $z$
 (namely $z=5, 10,$ and $20$, for $S=100$, and $z=50, 100,$ and $200$
 for $S=1000$). The thick solid line is the theoretical prediction in
 the limit $C \to 0$. One observes the collapse of distributions
 having the same $C$ but different $S$, similar to what we derived
 analytically for $C \to 0$. Note that the decay is faster than
 exponential for finite $C$.}
\label{fig1}
\end{figure}

In Figs.~\ref{fig1}(A) and (B) we compare the predictions of
Eq.~(\ref{prey}) with numerical simulations of the model. We find
close agreement between our analytical expression and the numerical
results. In particular, $p_{\rm prey}$ shows an exponential decay for
large $k$. The deviations observed for small values of $k$ are due to
the fact that $k_{j}=S a_{j}$ is a good approximation only when the
fluctuations of $k_{j}$ are small, which is not true for small $k$.

Equation (\ref{prey}) depends only on the scaled number of prey
$k/2z$. Thus, for {\it any} value of $z$, the scaled variable
$\tilde{k}=k/2z$ obeys the same probability density function,
\begin{equation}
\label{prey_pdf}
p_{\rm prey}(\tilde{k})= E_{1}(\tilde{k}) \,.
\end{equation}
This probability density function is therefore universal, i.e., it is
identical for any values of $S$ and $z$ provided that $C$ is
small. For finite $C$, $p_{\rm{prey}}$ has a truncation of the
exponential decay at a value of $\tilde{k}$ that is a decreasing
function of $C$. Remarkably, even for finite $C$, $p_{\rm{prey}}$ only
depends on $C$; cf. Fig.~\ref{fig2}(A).

\subsection{Distribution of number of predators}\label{secnpred}

For $C \ll 1$ the predators of species $i$ have niche values $n_j >
n_{i}$, and the segment $a_j$ is placed with equal probability in the
interval $[0, n_j]$. Therefore, the probability that a species $j$
preys on $i$, provided $n_j > n_{i}$, is
\begin{equation}
a_{j} / n_{j}= x_j n_j / n_j = x_j\,,
\end{equation}
implying that the average probability that a species with $n_j > n_i$
preys on species $i$ is $\overline{x}$.

\begin{figure}[t]
\centerline{\includegraphics*[width=\columnwidth]{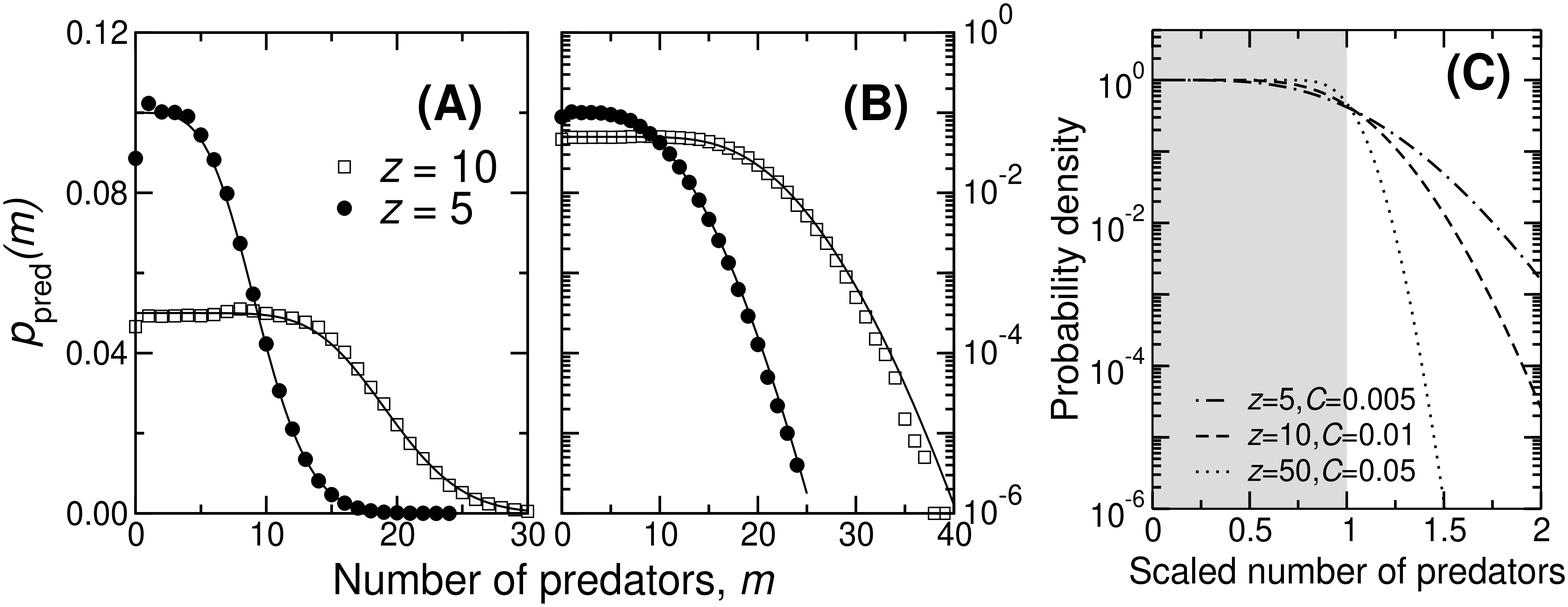}}
\caption{ (A) Linear and (B) log-linear plots of the distribution of
 number of predators for 1000 simulations of food webs with
 $S=1000$. We show results for $z=5$ and $10$ and the corresponding
 theoretical predictions. As predicted by Eq.~(\ref{pred}), we find a
 regime where the distribution is approximately uniform followed by a
 Gaussian decay. (C) Log-linear plot of the analytical expressions for
 the probability densities for the scaled number of predators $m/2z$
 for three values of $z$. For all values of $z$, we find an
 approximately constant value of $p_{\rm pred}$ for $m/2z < 1$ (shaded
 region) and a fast Gaussian-like decay for $m/2z > 1$.}
\label{fig2}
\end{figure}

If one notes that species $i$ has $(S-i)$ potential predators---those
species with $n_j > n_i$---then it follows that, in the limit $C \ll
1$, the total number of predators of $i$ is the result of $( S-i )$
independent ``coin-throws'' with probability $\overline{x}$ of a given
species $j$, with $n_j > n_i$, being a predator. This fact implies
that the probability of species $i$ having $m$ predators is given by
the binomial distribution,
\begin{equation}
p_{\rm pred}^{i}(m)={{S-i}\choose m} \overline{x}^{m}
(1-\overline{x})^{S-m-i}
\end{equation}
It then follows that the distribution of number of predators for the
set of all species in the web is an average over the binomials for the
different values of $i$
\begin{equation}
\label{predators}
p_{\rm pred}(m) = \frac{1}{S}\sum _{i=1}^{S-m}{p_{\rm
pred}^{i}(m)}\nonumber = \frac{1}{S}\sum _{i=1}^{S-m}{{S-i}\choose m}
\overline{x}^{m} (1-\overline{x})^{S-m-i}
\end{equation}
In the limit of interest, $S \gg 1$, and with $S \overline{x} = 2z$,
one can approximate the binomial distribution by a Poisson and the sum
by an integral, yielding
\begin{equation}
\label{pred}
p_{\rm pred}(m)=\frac{1}{2z}\int _{0}^{2z}dt\, \frac{t^{m}e^{-t}}{m!}=
\frac{1}{2z}~\gamma (m+1,2z),
\end{equation}
where $\gamma$ is the ``incomplete gamma function''
\citep{gradstheyn00}.
For $m < 2 z$, $p_{\rm pred}$ is approximately constant because
$\gamma (m+1,2z) \approx 1$, while for $m > 2z$ $p_{\rm pred}$ decays
with a Gaussian tail. In Figs.~\ref{fig1}(A) and (B), we compare the
predictions of Eq.~(\ref{pred}) with numerical simulations and find
good agreement.

Unlike the scaling seen for the distribution of number of prey,
Eq.~(\ref{pred}) is not simply a function of the scaled variable
$m/2z$. However, for small values of $m/2z$, $\gamma$ is a constant
and thus it does not depend on $m$ or $z$. So, the probability density
for the scaled variable $\tilde{m}=m/2z$,
\begin{equation}
p_{\rm pred}(\tilde{m})=\gamma(2z \,\tilde{m}+1,2 z) \approx 1
\quad\quad\quad \tilde{m}<1
\end{equation}
for {\it any} $z$. For $\tilde{m}>1$, $p_{\rm{pred}}(\tilde{m})$
decays as a Gaussian. In Fig.~\ref{fig2}(B) we show that the decay
rate of $p_{\rm pred}$ increases with $z$, becoming a sharp step drop
at $\tilde{m} = 1$ as $z \to \infty$.

\begin{figure}[t]
\centerline{\includegraphics*[height=0.5\columnwidth]{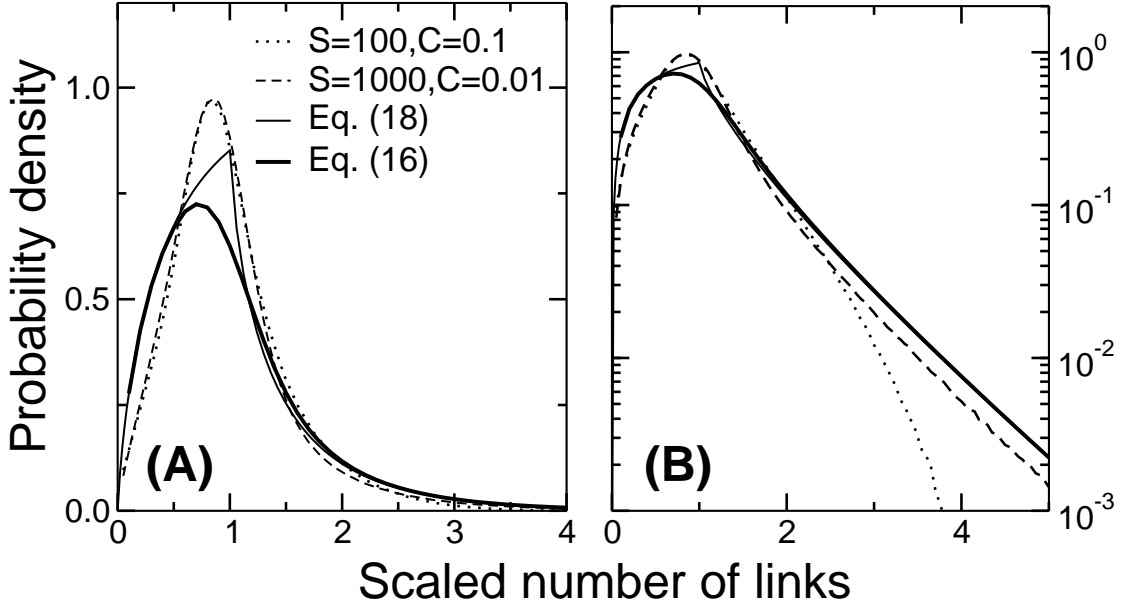}}
\caption{(A) Linear and (B) log-linear plots of the the probability
 densities for the scaled number of links, $\tilde{r}=r/2z$. We show
 results for $z=10$ and web sizes $S=100$ (dotted line) and $S=1000$
 (dashed line). The two curves show a similar behavior: A maximum
 around $\tilde{r}=1$, and a tail dominated by the distribution of
 number of prey. For $S=1000$, one has $C=0.01$ and the decay is
 exponential. For $S=100$, $C=0.1$ and the decay is faster than
 exponential (see Fig.~\protect{\ref{fig1}}). We also show two
 analytical curves in the limit $C \rightarrow 0$: the convolution,
 Eq.~(\ref{convol}), for $z=10$ (thick solid line) and for $z=\infty$
 (thin solid line). As with the simulations, the theoretical curves
 also display a maximum at $\tilde{r} \approx 1$. The decay for both
 analytical curves is exponential, as is the case for the distribution
 of number of prey for $C \rightarrow 0$.}
\label{fig3}
\end{figure}

\subsection{Distribution of number of trophic links}

The number $r$ of trophic links of a species is the sum of the number
of prey and number of predators of that species. In Fig.~\ref{fig3},
we display the probability densities $p_{\rm links}(\tilde{r}=r/2z)$
obtained from numerical simulations for $S=100$ and $S=1000$. Both
distributions show a maximum around $\tilde{r} = 1$ and the same decay
in the tail as the distribution of number of prey, i.e. an exponential
decay for small $C$, and faster than exponential decay for larger $C$
values. This similarity is expected since the distribution of number
of predators decays very rapidly, so that large values of $r$ must
arise due to large values of $k$.

Obtaining an analytical solution for $p_{\rm links}(r)$ is
considerably more difficult than for $p_{\rm prey}$ or $p_{\rm pred}$
because it requires one to estimate the correlations between the
number of prey and the number of predators. The rules underlying the
niche model imply that the number of predators is negatively
correlated with the number of prey. That is to say, a species having a
large number of predators is likely to have a low niche number, and
thus it will likely predate upon few or no species. This attribute of
the niche model is discussed in more detail in Section 2.7.


If one neglects, for the moment, correlations between the distribution
of number of prey and number of predators, the probability density for
the scaled total number of links, $\tilde{r}=r/2z$, is found simply by
the convolution of both distributions. In the limit $C \ll 1$, one has
\begin{eqnarray}
\label{convol}
p_{\rm links}(\tilde{r}) &=& \int_0^{\tilde{r}} p_{\rm{prey}}(t)
 p_{\rm{pred}}(\tilde{r}-t) \,dt\nonumber\\ &=& \int_0^{\tilde{r}} E_{1}(t)
 \gamma(2z(\tilde{r}-t)+1,2z) \,dt
\end{eqnarray}

This integral cannot be expressed in terms of recognizable functions,
but can be calculated numerically. In Fig.~\ref{fig3} we plot the
prediction of Eq.~(\ref{convol}) for $z=10$. As we obtained for the
simulations of the model, we find a maximum of $p_{\rm links}$ around
$\tilde{r}=1$ and an exponential decay in the tail. The latter arises
from the exponential tail found for the distribution of number of
prey, as previously mentioned.

One can also obtain an explicit expression for $p_{\rm links}$ when $z
\to \infty$. One expects this solution not to differ much from that
for a finite $z$, since the only difference is that a smooth step in
the distribution of the number of predators around $\tilde{m}=1$ is
substituted by a sharp one. As discussed in the previous section, the
limits of $z \to \infty$ and $C \ll 1$ yield
\begin{eqnarray}
p_{\rm pred}(m)=\left\{ \begin{array}
{l@{\quad}l}
1 & \mbox{if}\quad m<2 z,\\
0 & \mbox{if}\quad m>2 z.
\end{array} \right.
\label{ppredsharp}
\end{eqnarray}
Integration of Eq.~(\ref{convol}) then yields
\begin{eqnarray}
 p_{\rm links}(\tilde{r})=\left\{ \begin{array}{l@{\quad \quad}l}
  1 - E_{2}(\tilde{r}) & \mbox{if}\quad \tilde{r}<1,\\
  E_{2}(\tilde{r}-1)-E_{2}(\tilde{r}) & \mbox{if}\quad \tilde{r}>1.
\end{array} \right.
\label{convol2}
\end{eqnarray}
with $E_{2}(x)=1 - exp(-x) + x E_{1}(x)$ the exponential-integral of
order 2 \citep{gradstheyn00}. Figure \ref{fig3} also displays
Eq.~(\ref{convol2}), showing a curve rather similar to the convolution
for $z=10$ except close to $\tilde{r}=1$.

From the analytical solution, one can demonstrate that the tail of
$p_{\rm links}$ decays exponentially. For large $\tilde{r}$,
Eq.~(\ref{convol2}) can be approximated by
\begin{eqnarray}
\label{expintexpan}
E_{2}(\tilde{r}-1)-E_{2}(\tilde{r}) &\simeq& - E_{2}'(\tilde{r}-1/2)\nonumber
\\&=& ~~ E_{1}(\tilde{r}-1/2)\nonumber\\
&\simeq& ~~\frac{exp(-\tilde{r}+1/2)}{\tilde{r}-1/2}\,,
\end{eqnarray}
where $E_{2}'=-E_{1}$ is the derivative of function $E_{2}$ and the last
expression is the dominant term in the expansion of
exponential-integral functions \citep{gradstheyn00}. Note that
Eq.~(\ref{expintexpan}) is a good approximation of Eq.~(\ref{convol2})
for $\tilde{r} > 2$.


\subsection{Fraction of top, intermediate, basal and cannibal species}\label{seccan}

%
\begin{figure}[t]
 \vspace*{0.cm}
\centerline{\includegraphics*[width=0.8\columnwidth]{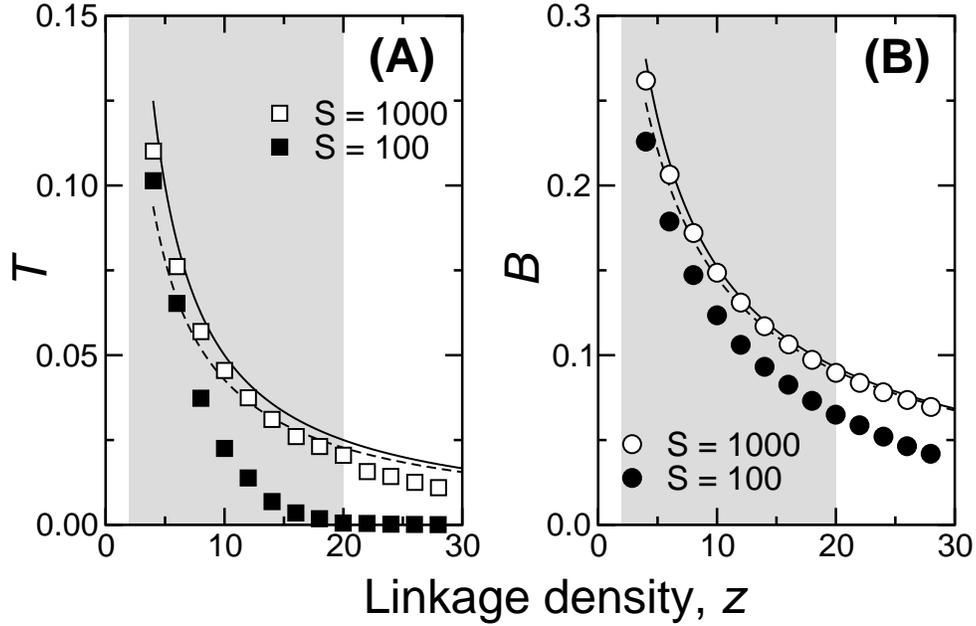}}
 \vspace*{-0.3cm}
 \caption{Fraction of top and basal species as a function of the
 linkage density $z$. The shaded region corresponds to the interval
 of $z$ typically observed in empirical food webs
 \protect\citep{dunne02}. (A) Comparison of the results for 1000
 simulations of food webs for which isolated species were removed
 with our theoretical prediction Eqs.~(\protect\ref{top}), solid
 line, and (\protect\ref{tb1}), dashed line, for fraction of top
 species. (B) Comparison of the results of 1000 simulations of food
 webs for which isolated species were removed with our theoretical
 prediction Eqs.~(\protect\ref{basal}), solid line, and
 (\protect\ref{tb1}), dashed line, for fraction of basal
 species. Note that for both $T$ and $B$ there is good agreement
 between the analytical expressions and the numerical results when $C
 \ll 1$. Also note that the theoretical predictions provide narrow
 bounds for the numerical results in this limit.}
 \label{fig4}
\end{figure}

As the names indicate, top species have no predators and basal species
have no prey, while intermediate species are those with both prey and
predators. The fraction $T$ of top species is, by definition,
\begin{equation}
\label{top}
T \equiv p_{\rm pred}(0) =\frac{1-\exp(-2z)}{2z}\,,
\end{equation}
as directly seen from Eq.~(\ref{pred}). Typically empirical webs have
$2 < 2z < 20$, a region for which Eq.~(\ref{top}) can be approximated
simply as $T=1/2z$.

%

To calculate the fraction $B$ of basal species, we note that a species
has no prey only if its range $a$ falls in a region with no
species. In the limit of large $S$, the probability density for
finding an empty interval of length $\delta$ is $S e^{-S\delta}$, as
predicted by the canonical distribution \citep{pathria72}. Thus, the
probability of finding a species-free segment of length larger than
$a$ is $e^{-Sa}$, which yields the probability for a species with a range
$a$ not to prey on other species. Using Eq.~(\ref{presas}), it follows
that the fraction of bottom species is:
\begin{equation}
\label{basal}
B = \int _{0}^{1}da\, e^{-Sa}p(a) = \frac{\ln(1+2z)}{2z}.
\end{equation}
Note that we must take this approach to the calculation because the
definition of the fraction $B$ of basal species, $p_{\rm prey}(0)$,
leads, in the continuous analysis, to a divergence as the
exponential-integral function $E_{1}(x)$ diverges in the limit $x\to
0$.
%

%
In the niche model, isolated species are eliminated, so they are not
counted toward top and basal species. To correct Eqs.~(\ref{top}) and
(\ref{basal}) for this effect, we must account for the isolated
species in our calculations. We estimate the number of isolated
species to first order by assuming that having no prey is
statistically independent of having no predators, implying that the
fraction of isolated species is just the product of the fractions of
top and basal species. This assumption does not take into account the
(negative) correlations between the number of prey of a species and
its number of predators. Nonetheless, this simple approximation
provides an upper bound for the number of isolated species which leads
to lower bounds on $T$ and $B$,
\begin{equation}
\label{tb1}
T'=\frac{T-TB}{1-TB}\,, \quad\quad\quad B'=\frac{B-TB}{1-TB}
\end{equation}
In Fig.~\ref{fig4}, we compare our analytical predictions for the
fraction of top and basal species with numerical simulations of the
model. As expected, Eqs.~(\ref{top}--\ref{tb1}) provide bounds for the
numerical results in the limit $C \ll 1$.

The fraction of intermediate species is just $I = 1 - (T + B)$.

\begin{figure}[t!]
 \vspace*{0.cm}
 \centerline{\includegraphics*[width=0.5\columnwidth]{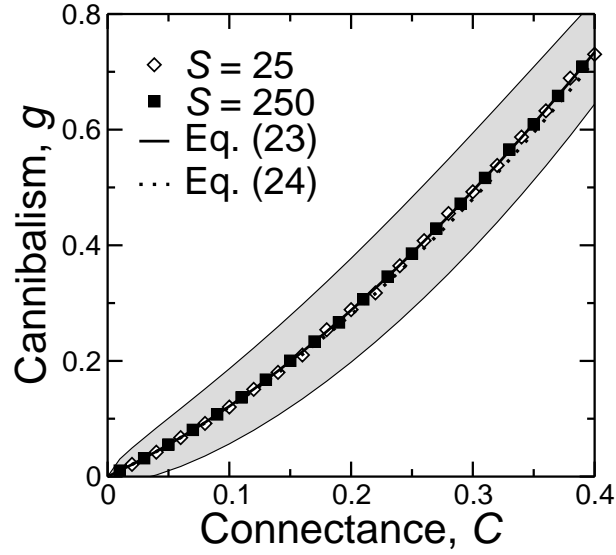}}
 \vspace*{-0.3cm}
 \caption{Fraction of cannibals as a function of the connectance
 $C$. We plot the fraction of cannibal species averaged over 5000
 realizations of the model for two system sizes, and the predictions
 of Eqs.~(\ref{can1}) and (\ref{can2}). It is visually apparent that
 the results are insensitive to changes in $S$ and that they are in
 good agreement with our theoretical predictions. The shaded area
 indicates one standard deviation from the average for $S=25$; the
 upper and lower curves have been obtained using Eqs.~(\ref{deltag})
 and (\ref{can2}). Our results demonstrate that in the niche model the
 fraction of cannibals depends only upon the
 connectance.\vspace{0cm}}
 \label{fig6}
\end{figure}

Another quantity of ecological interest is the fraction of cannibal
species. Consider a species $i$ with niche number $n_i$. Its preys are
the species inside an interval of length $a_i=x n_i$, chosen according
to Eq.~(\ref{px}), and centered with uniform probability in the
interval $[a_{i}/2,n_{i}]$. For $i$ to predate on itself, it is
necessary that the segment of length $a_i$ contains species $i$. This
is equivalent to the distance between the center of the segment and
$n_i$ being smaller than $a_{i}/2$, which occurs with probability
$\frac{a_i}{2}/(n_i-\frac{a_i}{2})$. Therefore, the average
probability $g$ for a species to be a cannibal is given by
\begin{equation}
g = \int_{0}^{1} b (1-x)^{b-1} \frac{x/2}{1-x/2} dx \label{can0} =
\,_{2}F_{1}(1,1;2C;1/2) -1
\label{can1}
\end{equation}
where $_{2}F_{1}$ is the Gauss
hypergeometric function \citep{gradstheyn00}. One thus finds that, for
any value of $S$, the fraction of cannibals depends only upon
$C$. Equation (\ref{can1}) can be expanded for small $C$ as
\begin{equation}
 g \simeq C+2 C^2 + O(C^3).
\label{can2}
\end{equation}
Since the largest value of $C$ reported in the literature is $0.3$, we
expect Eq.~(\ref{can2}) to provide a good approximation for
empirically relevant values of $C$ (Fig.~\ref{fig6}).

Next, we evaluate the standard deviation of $g$. If $\ell$ denotes the
number of cannibal species in a single realization of the model with
$S$ species, the corresponding fraction of cannibals is simply
$g_{\ell}=\ell/S$. Therefore, the standard deviation of $g_{\ell}$ is
$\sigma_{g}=\Delta \ell/S$. Since $g$ is the probabilitiy for a species
to be a cannibal, the probability of having exactly $\ell$ cannibal
species is given by a binomial distribution
\begin{equation}
p_{can}(\ell)={{S}\choose \ell} g^{\ell} (1-g)^{S-\ell}.
\end{equation}
Therefore, $\Delta \ell^{2}= Sg(1-g)$, and the standard deviation for the
fraction of cannibalism is
\begin{equation}
\sigma_{g}=\sqrt{\frac{g(1-g)}{S}}\,.
\label{deltag}
\end{equation}
This expression implies that $\sigma_g$ decreases with increasing $S$.


\subsection{Standard deviation of generality and vulnerability}

\begin{figure}[t]
 \centerline{\includegraphics*[width=0.8\columnwidth]{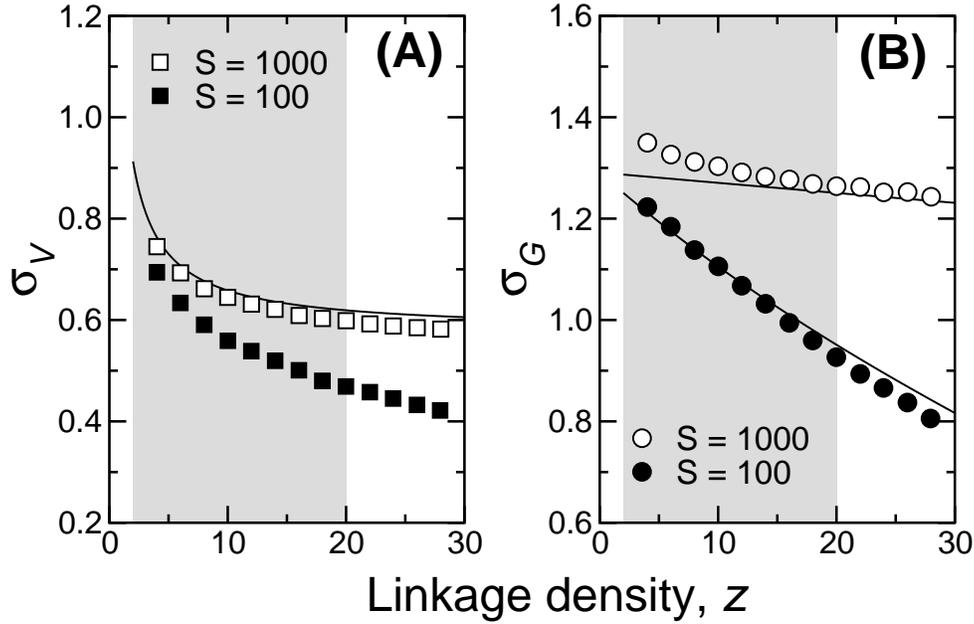}}
 \caption{Normalized standard deviations of generality and
 vulnerability as a function of the average linkage density $z$. The
 shaded region corresponds to the interval of $z$ typically observed
 in empirical food webs \protect\citep{dunne02}. (A) Comparison of the
 results for 1000 simulations of food webs for which isolated species
 were removed with our theoretical prediction
 Eq.~(\protect\ref{deltaV}) for the standard deviation of the
 vulnerability. (B) Comparison of the results for 1000 simulations of
 food webs for which isolated species where removed with our
 theoretical prediction Eq.~(\protect\ref{deltaG}) for the standard
 deviation of the generality. Note that Eq.~(\protect\ref{deltaV}) is
 only valid in the limits of $S \gg 1$ and $C \ll 1$, which is
 confirmed in (A) by the deviations found for $S=100$. Equation
 (\protect\ref{deltaG}) was derived independent of these limits and
 thus holds for any value of $C$. The small underestimation
 of our expression for $\sigma_{G}$ relates to the fact that $k_j = S
 r_j$ is a good approximation only when the fluctuations of $k_j$ are
 small, which is no longer true for small $k$.\vspace{0cm}}
 \label{fig5}
\end{figure}

The vulnerability of a prey is defined as its number $m$ of predators,
and the generality of a predator as its number $k$ of prey. Following
Williams and Martinez (\citeyear{martinez00}), we define the
normalized standard deviations of the vulnerability as $\sigma_V^2=
\overline{m^{2}}/\overline{m}^{2}-1$ and of the generality as
$\sigma_G^2= \overline{k^{2}}/\overline{k}^{2}-1$. By definition, one
has $\overline{m}=\overline{k}=z$.

To evaluate $\sigma_V$, we first calculate
$\overline{m^{2}}$. Equation~(\ref{pred}) yields $\overline{m^{2}}=
4z^2/3 + z$, so that
\begin{equation}
\sigma_{V}=\sqrt{\frac{1}{3}+\frac{1}{z}}\,,
\label{deltaV}
\end{equation}
which approaches $\sqrt{1/3}$ as $z \to \infty$
(Fig.~\ref{fig5}). Since Eq.~(\ref{pred}) is valid only in the limits
$S \gg 1$ and $C \ll 1$, these are also the limits of validity for
Eq.~(\ref{deltaV}). We present these results, as well as simulations
of the niche model in Fig.~\ref{fig5}(A).

We next calculate $\sigma_G$ by direct evaluation of
$\overline{k^{2}}$. If $S \gg 1$, the number of prey of a species
having a range $a$ is $k=S a$, yielding
\begin{equation}
\frac{\overline{k^{2}}}{\overline{k}^{2}} =
\frac{\overline{a^{2}}}{\overline{a}^{2}} =\frac{\overline{n^{2}~ x^{2}}}
{\overline{n~x}^{2}} = \frac{8(b+1)}{3(b+2)}.
\end{equation}
Thus, $\sigma_G$ is
\begin{equation}
\sigma_{G}=\sqrt{\frac{8}{3} \frac{1}{1+2 C}-1}.
\label{deltaG}
\end{equation}
For $C \ll 1$, $\sigma_G = \sqrt{5/3}$, a result that can also be
obtained from Eq.~(\ref{prey}). Unlike Eq.~(\ref{deltaV}),
Eq.~(\ref{deltaG}) depends upon $C$ and thus does not require the same
limits considered in the derivation of Eq.~(\ref{deltaV}). We show
these results as well as those for numerical simulations of the niche
model in Fig.~\ref{fig5}(B).

\subsection{Correlation coefficient}\label{seccor}

The mechanisms by which webs are constructed in the niche model imply
that a species with many predators is likely to have a low niche
value. Furthermore, a species with a low niche value probably feeds on
few prey. As a consequence, the more predators a species has, the
smaller the number of prey will be, and vice versa. Thus, the number
of prey of a species and its number of predators must be negatively
correlated.

\begin{figure}[t]
 \vspace*{0.cm}
\centerline{\includegraphics*[width=0.8\columnwidth]{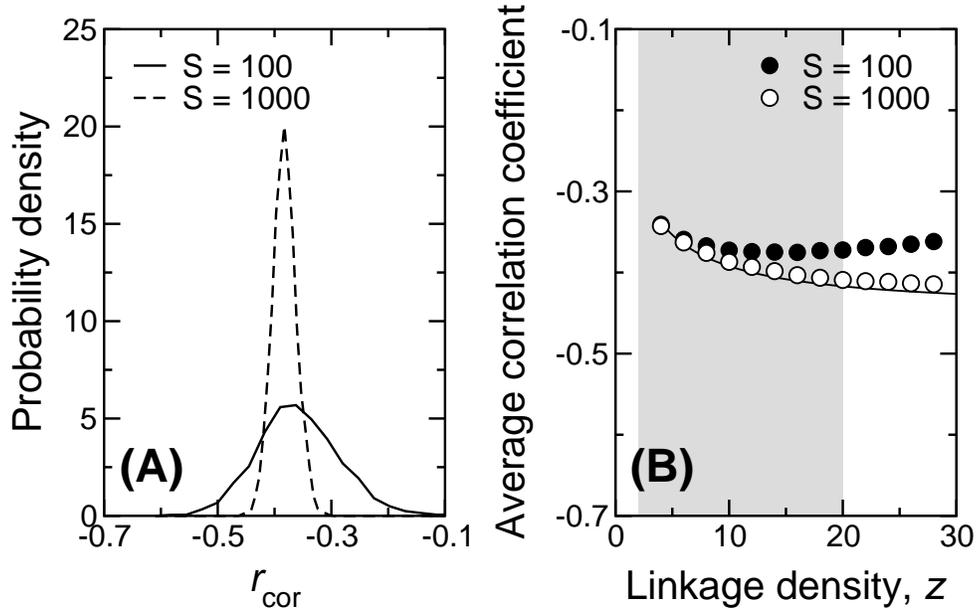}}
 \vspace*{-0.3cm}
 \caption{(A) Distribution of $r_{\rm cor}$, the correlation
 coefficient between number of prey and number of predators obtained
 from 5000 realizations of the niche model. We show simulation results
 for $z=10$. (B) Comparison of the average correlation coefficient
 obtained from 5000 simulations of the model (for which isolated
 species where removed) with our theoretical prediction,
 Eq.~(\protect\ref{expcorr}). The negative correlations observed are
 inherent to the model, as a species having a large number of
 predators is likely to have a low niche number, and consequently will
 predate upon few or no species.}
 \label{modelcorrelation}
\end{figure}

In order to quantify the correlations between $k$ and $m$ we define
the correlation coefficient between the number of prey and the number
of predators,
\begin{equation}
r_{\rm cor}=\frac{\overline{km}-\overline{k}\,\overline{m}}
{\sqrt{\overline{k^{2}}-{\overline{k}}^{2}}\,
\sqrt{\overline{m^{2}}-{\overline{m}}^{2}}\,}
\label{corr}
\end{equation}
where, as before, the bar over a variable denotes an average over all
species in a web.

We evaluate the average correlation coefficient, $r_{\rm cor}$, for
the niche model in the limits $S \gg 1$ and $C \ll 1$. Equation
(\ref{corr}) can be rearranged to yield
\begin{equation}
r_{\rm cor}=\frac{\overline{k m}/z^{2}-1}{\sigma_{G}\sigma_{V}}\,,
\label{corr2}
\end{equation}
where $\sigma_{G}$ and $\sigma_{V}$ are the normalized standard
deviations of the generality and the vulnerability respectively. Thus,
in order to obtain a closed form for Eq.~(\ref{corr2}), we need only
to calculate
\begin{equation}
\overline{km} \equiv \sum_{k,m=0}^{S} k m~p(k,m)\,,
\label{km}
\end{equation}
where $p(k,m)$ is the probability that a species has $k$ prey and $m$
predators. If $k$ and $m$ were independent, $p(k,m)$ would be simply
the product $p_{\rm prey}(k) p_{\rm pred}(m)$, yielding
\begin{equation}
\overline{km} = \sum_{k=0}^{S}{k p_{\rm prey}(k)} \sum_{m=0}^{S}{m
p_{\rm pred}(m)}=\overline{k} \overline{m} = z^2
\end{equation}
and $r_{\rm corr}=0$. However, $k$ and $m$ are not independent.

Let us define $p(k,m,n)$ as the probability that a species has $k$
prey, $m$ predators, and niche number $n$. One then has
\begin{equation} 
p(k,m,n)= p(k|m,n) p(m|n) p_{n}(n) \,.
\end{equation}
where $p(~|~)$ refers to the conditional probability.

Remembering that $p_{n}(n)=1$ and noticing that $p(k|m,n)=p(k|n)$
because the number of prey in the niche model is determined solely by
the niche value $n$, one can write
\begin{equation}
p(k,m)\equiv \int dn \, p(k,m,n) = \int dn \, p(k|n) p(m|n) \,.
\label{pkm}
\end{equation}

Using Eq.~(\ref{pkm}) in Eq.~(\ref{km}) and rearranging terms, yields
\begin{equation}
\overline{km}= \int_{0}^{1}dn \, \overline{k_{n}} \, \overline{m_{n}}
\,,
\label{km2}
\end{equation}
with $\overline{k_{n}}$ and $\overline{m_{n}}$ the average number of
prey and of predators for a species with niche number $n$. One can
calculate $\overline{k_{n}}$ by realizing that $k_{n}=S n x$ and $S
\overline{x} = 2z$, so that
\begin{equation}
\overline{k_{n}}=S n \overline{x} = 2 z n \,.
\label{kofn}
\end{equation}
In the limits $S \gg 1$ and $C \ll 1$, the potential predators of a
species with niche value $n$ are those species with niche values
larger than $n$. Each of these species has a probability
$\overline{x}$ of preying on it. Therefore, the average number of
predators for a species with niche value $n$ is
\begin{equation}
\overline{m_{n}}= S (1-n) \overline{x} = 2 z (1-n) \,. \nonumber
\label{mofn}
\end{equation}
Substituting Eqs.~(\ref{kofn}) and (\ref{mofn}) into Eq.~(\ref{km2}),
one obtains $\overline{km}=2 z^{2}/3$. Finally, using
Eq.(\ref{deltaV}) and Eq.(\ref{deltaG}) in the limit $C \rightarrow 0$
one obtains
\begin{equation}
r_{\rm cor}=-\frac{1}{\sqrt{5(1+3/z)}} \,.
\label{expcorr}
\end{equation}

In Fig.~\ref{modelcorrelation}(A), we show the probability density of
model realized $r_{\rm cor}$. The figure demonstrates that
fluctuations from the average value are not negligible for $S$ as
large as 100. In Fig.~\ref{modelcorrelation}(B) we compare
Eq.(\ref{expcorr}) with numerical simulations of the niche model,
finding good agreement in the limits of interest, $S \gg 1$ and $C \ll
1$. The two figures confirm our initial hypothesis that the niche
model generates webs with negative correlations between $k$ and $m$.

\subsection{Assortativity}\label{secassor}

The assortativity of a food web quantifies the correlation between the
number of links---trophic interactions---of a species and the number
of links of its prey and predators \citep{newman02}. In a food web
with positive assortativity---an assortative food web---highly
connected species tend to be connected to one another. Conversely, in
a food web with negative assortativity---a disassortative food
web---highly connected species tend to connect to species with a small
number of links. Prior work by \cite{newman02} suggests that
assortativity tends to be positive for social networks and negative
for technological and biological networks. In this subsection we
evaluate the assortativity of the food webs generated by the niche
model in the limit of large $S$ and small $C$.

Let us consider a trophic interaction in the network, in which the
predator has $r$ trophic links and the prey has $r'$ trophic
links. The assortativity $A$ is defined as \citep{newman02}
\begin{equation}
A=\frac{\langle r r'\rangle -\langle \frac{r+r'}{2}\rangle ^{2}}{\langle \frac{r^{2}+r'^{2}}{2}\rangle -\langle \frac{r+r'}{2}\rangle ^{2}},
\label{assort}
\end{equation}
where $\langle\dots\rangle$ denotes the average over all trophic links
in the food web.

We first calculate $\langle r r'\rangle $. In section \ref{seccor} we
obtained the average number of prey and the average number of
predators for a species with niche number $n$, namely
$\overline{k_{n}}= 2 z n$, and $\overline{m_{n}}= 2 z (1-n)$. Thus the
average number of links of a species with niche number $n$ is
$\overline{r_{n}}= \overline{k_{n}}+ \overline{m_{n}}= 2 z$, which is
independent of $n$. It follows then that the average number of links
is the same for all species, independently of the niche
number. Therefore, $r$ and $r'$ are independent variables, so that
$\langle r r'\rangle = \langle r\rangle \langle r'\rangle $.

\begin{figure}[t]
 \vspace*{0.cm}
\centerline{\includegraphics*[width=0.8\columnwidth]{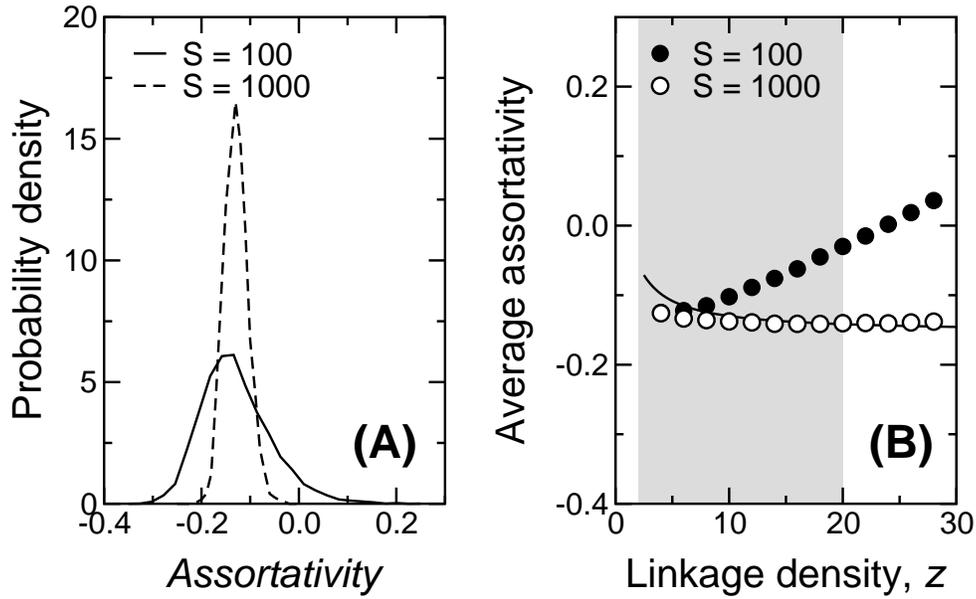}}
 \vspace*{-0.3cm}
 \caption{(A) Distribution of the assortativity $A$ obtained from 5000
 realizations of the niche model. We show simulation results for
 $z=10$. (B) Comparison of the average assortativity obtained from
 5000 simulations of the model (for which isolated species where
 removed) with our theoretical prediction,
 Eq.~(\protect\ref{aniche}). As predicted by
 Eq.~(\protect\ref{aniche}), the niche model yields webs that are
 slightly disassortative in the limit of large $S$. For small $S$,
 food webs become slightly assortative at large linkage densities.}
 \label{modelassortativity}
\end{figure}

In order to evaluate these averages, we transform the averages over
links, $\langle\cdots\rangle $, to averages over species,
$\overline{\cdots}$. Let us call $r_{i}$ the number of links of the
predator in link $i$, by definition $\langle r\rangle $ is
\begin{equation}
\langle r\rangle =\frac{\sum_{\rm links}r_{i}}{L} \,,
\end{equation}
where $L$ is the number of links in the network. The predator from
link $i$---denoted as species $j$---has $k_{j}$ prey, so that it
appears as the predator in $k_{j}$ links. Therefore, one can write
\begin{equation}
\langle r\rangle =\frac{\sum_{\rm species}k_{j}\, r_{j}}{L}\,.
\end{equation}
Since $r_{j}=k_{j}+m_{j}$, the numerator can be rewritten as
$S(\overline{k^{2}}+\overline{k \, m})$, with the overbar denoting an
average over species, as used in previous sections. Finally, one has
\begin{equation}
\langle r\rangle =\frac{\overline{k^{2}}+\overline{k \, m}}{z} \,,
\label{rr}
\end{equation}
Analogously, one can obtain
\begin{equation}
\langle r'\rangle =\frac{\overline{m^{2}}+\overline{k \, m}}{z} \,,
\label{r'}
\end{equation}
\begin{equation}
\langle r^{2}\rangle =\frac{\overline{k^{3}}+\overline{k \, m^{2}}+ 3\, \overline{k^{2} m}}{z} \,,
\label{r2}
\end{equation}
\begin{equation}
\langle r'^{2}\rangle =\frac{\overline{m^{3}}+\overline{k^{2} m}+ 3 \,\overline{k \, m^{2}}}{z} \,.
\label{r'2}
\end{equation}
These averages can be evaluated as in section \ref{seccor} (see
Appendix A). Substituting Eqs.~(\ref{rr})--(\ref{r'2}) into
Eq.~(\ref{assort}), one obtains
\begin{equation}
A= -\frac{(2z/3 -1/2)^{2}}{26z^{2}/9+z/3+1/4} \le 0\,.
\label{aniche}
\end{equation}
In Fig.~\ref{modelassortativity}, we compare this expression with
numerical simulations and find good agreement in the limits for which
our derivation is valid.


\subsection{Robustness of the analytical results}

In Table~\ref{expressions} we summarize the analytical results
obtained for the niche model. An important question is that of the
robustness of our predictions to changes in the particular formulation
of the {\it details} of the model. The approximations used in the
derivations of the expressions for the distributions of the number of
prey and predators, Eqs.~(\ref{prey}) and (\ref{pred}), allow us to
conclude that:
\begin{enumerate}
\item The distribution of number of prey depends on the functional
form of $p(x)$, but Eq.~(\ref{prey}) will still be obtained for any
$p(x)$ decaying exponentially as $C$ (or $\overline{x}$) tends to
zero.
\item The distribution of number of predators does not depend on the
specific form of $p(x)$. The only requirement is that the connectance
$C=\overline{x}/2$ tends to zero under some limit, so that $2z=S C$
remains finite when $S \to \infty$.
\end{enumerate}
\begin{table}
\begin{tabular}{lc}
\hline
Property & Expression \\
\hline
Distribution of number of prey & $p_{\rm prey}(k)=(1/2z)\,E_1(k/2z)$ \\
Distribution of number of predators & $p_{\rm pred}(m)=(1/2z)\,\gamma(m+1,2z)$\\
Fraction of top species & $T=\frac{1-\exp{(-2z)}}{2z}$ \\
Fraction of basal species & $B=\frac{\ln{(1+2z)}}{2z}$ \\
Fraction of cannibals & $g=\,_{2}F_{1}(1,1;2C;1/2)-1$\\
 & $\approx C+2C^2+O(C^3)$\\
Standard deviation of the vulnerability& $\sigma_V=\sqrt{1/3+1/z}$ \\
Standard deviation of the generality & $\sigma_G=\sqrt{8/(3+6C)-1}$ \\
Prey-predators correlation coefficient & $r_{\rm cor}=-\frac{1}{\sqrt{5(1+3/z)}}$\\
Assortativity & $A=-\frac{(2z/3-1/2)^2}{26z^2/9+z/3+1/4}$\\
\hline
\end{tabular}
\vspace{0.5cm}
\caption{Summary of the analytical expressions obtained for the niche
 model in the limits $S\gg1$ and $C\ll1$.}
\label{expressions}
\end{table}

Thus, it is safe to conclude that our findings are robust under quite
general conditions, a result that is not possible to obtain without an
analytic treatment of the problem. Moreover, as we will show in the
next section, the quantitative patterns uncovered for the niche model
(Table \ref{expressions}) are important guides for the study of
empirical food webs.

%
\section{Patterns in food web structure}

In this section, we compare the predictions of our theoretical
solution of the niche model with data from empirical food webs
obtained from a variety of environments. Remarkably, we find that the
quantitative patterns uncovered for the niche model also describe the
properties of food webs pertaining to very diverse habitats, such as
freshwater, marine-freshwater interfaces, and terrestrial
ecosystems.

\subsection{Cumulative distributions}

The reported empirical food webs generally contain a small number of
trophic species. This fact implies that the empirical distributions of
the number of prey and number of predators will be quite noisy. For
this reason, we consider here the cumulative distributions instead of
probability density functions.

We first derive an analytical expression for the cumulative
distribution of number of prey, $P_{{\rm prey}}(>k) = \sum_{k^{\prime}
\ge k} p_{{\rm prey}}(k^{\prime})$, in the limit of large sizes and
small connectances. Since $1/2z$ is generally quite smaller than $1$,
one can substitute the sum by an integral with negligible error,
yielding
\begin{eqnarray}
P_{{\rm prey}}(>k) &=& E_2\left(\frac{k}{2z} \right)\nonumber\\
&=&\exp\left( -\frac{k}{2z} \right) - \frac{k}{2z}%
\,E_1\left(\frac{k}{2z} \right) \,,
\label{cumprey}
\end{eqnarray}
which, similarly to Eq.~(\ref{prey}), predicts an exponential decay
for large $k$. In terms of the scaled variable $\tilde{k}=k/2z$, we
obtain
\begin{equation}
P_{{\rm prey}}(>\tilde{k})= \exp\left( -\tilde{k} \right) - \tilde{k}
\,E_1\left(\tilde{k} \right) \,,
\label{cumprey2}
\end{equation}
which, as Eq.~(\ref{prey}), contains {\it no free parameters}.

Next, we derive an analytical expression for the cumulative
distribution of number of predators. In the limits $S \gg 1$ and $C
\ll 1$, one can use Eq.~(\ref{pred}) to obtain
\begin{equation}
P_{{\rm pred}}(>m) = \frac{1}{2z} ~\sum_{m^{\prime}=m}^\infty ~\gamma
(m^{\prime}+1,2z) \,, 
\label{predcum}
\end{equation}
%
As we have already noted, for $m<2z$ the gamma function can be
approximated as $\gamma(m+1,2z) \simeq 1$. We can then rewrite
Eq.~(\ref{predcum}) as
\begin{equation}
\label{predcum2}
P_{{\rm pred}}(>m)=1-\frac{1}{2z} ~\sum_{m'=0}^{m-1}~\gamma(m'+1,2z) \simeq
1-\frac{m}{2z}\quad,\quad m<2z\,.
\end{equation}
For $m \ge 2z$, Eq.~(\ref{predcum2}) decays as the error
function \citep{gradstheyn00}.

\begin{figure}[p!]
 \vspace*{0.cm}
 \centerline{\includegraphics*[width=0.65\columnwidth]{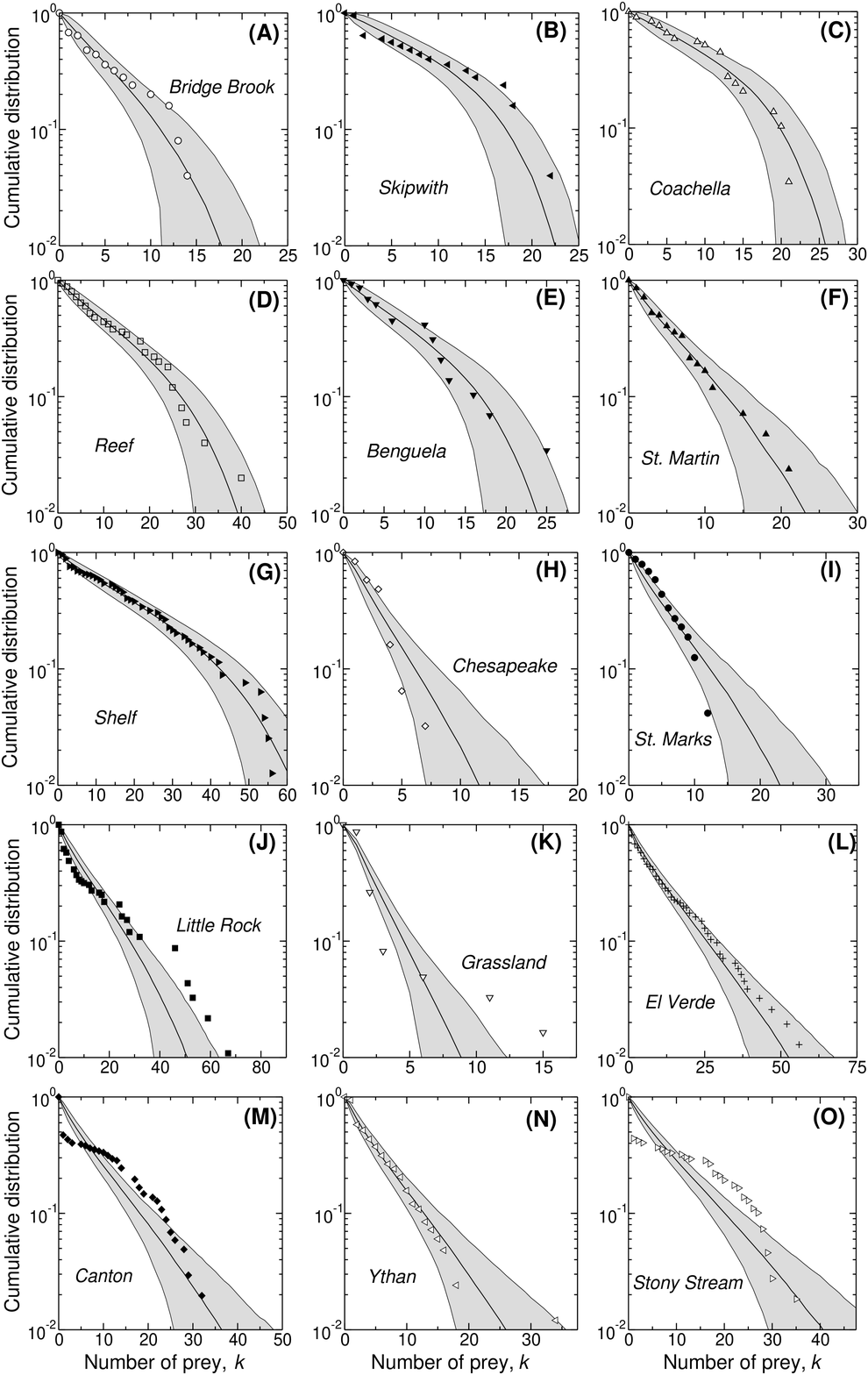}}
 \vspace*{-0.3cm}
 \caption{ Cumulative distribution $P_{{\rm prey}}$ of number $k$ of
 prey for the fifteen food webs studied: Bridge Brook Lake
 \protect\citep{havens92}; Skipwith Pond \protect\citep{warren89};
 Coachella Valley \protect\citep{polis91}; Caribbean Reef
 \protect\citep{opitz96}; Benguela \protect\citep{yodzis98};
 St. Martin Island \protect\citep{goldwasser93}; Northeast US Shelf
 \protect\citep{link02}; Chesapeake Bay \protect\citep{baird89};
 St. Marks Seagrass \protect\citep{christian99}; Little Rock Lake
 \protect\citep{martinez91}; Grassland \protect\citep{martinez99}; El
 Verde Rainforest \protect\citep{waide96}; Canton Creek
 \protect\citep{townsend98}; Ythan Estuary \protect\citep{hall91}; and
 Stony Stream \protect\citep{townsend98}. The solid black line
 represents the average value from 1000 simulations of the niche model
 and the grey region represents two standard deviations above and
 below the model's predictions.}
 \label{preypanel}
\end{figure}

\begin{figure}[t!]
 \vspace*{0.cm}
 \centerline{\includegraphics*[width=0.65\columnwidth]{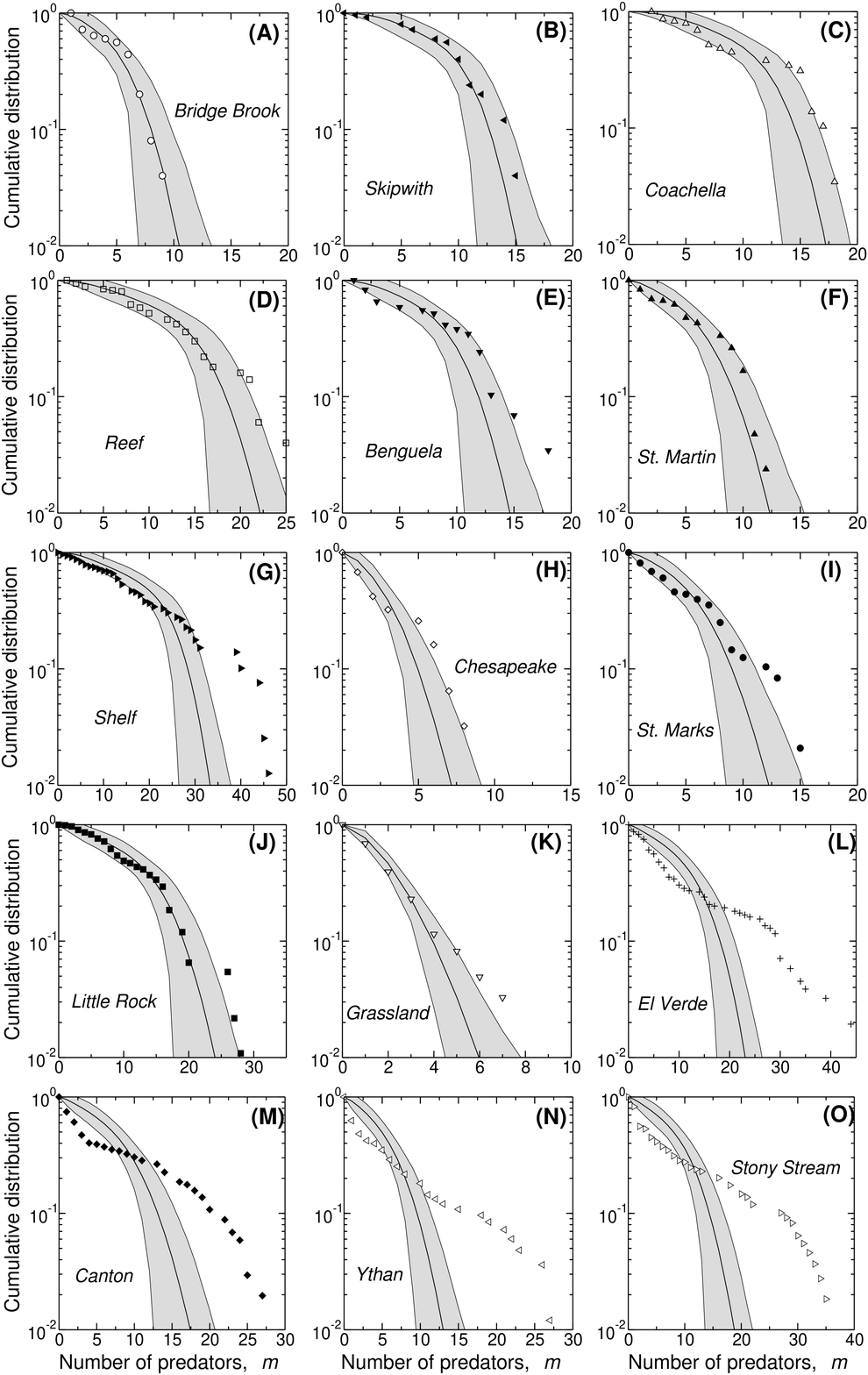}}
 \vspace*{-0.3cm}
 \caption{ Cumulative distribution $P_{{\rm pred}}$ of number $m$ of
 predators for the fifteen food webs studied (see
 Fig.~\protect\ref{preypanel}). The solid black line represents the
 average value from 1000 simulations of the niche model and
 the grey region represents two standard deviations above and below
 the model's predictions.\vspace{0cm}}
 \label{predpanel}
\end{figure}

\begin{figure}[t!]
 \vspace*{0.cm}
 \centerline{\includegraphics*[width=0.65\columnwidth]{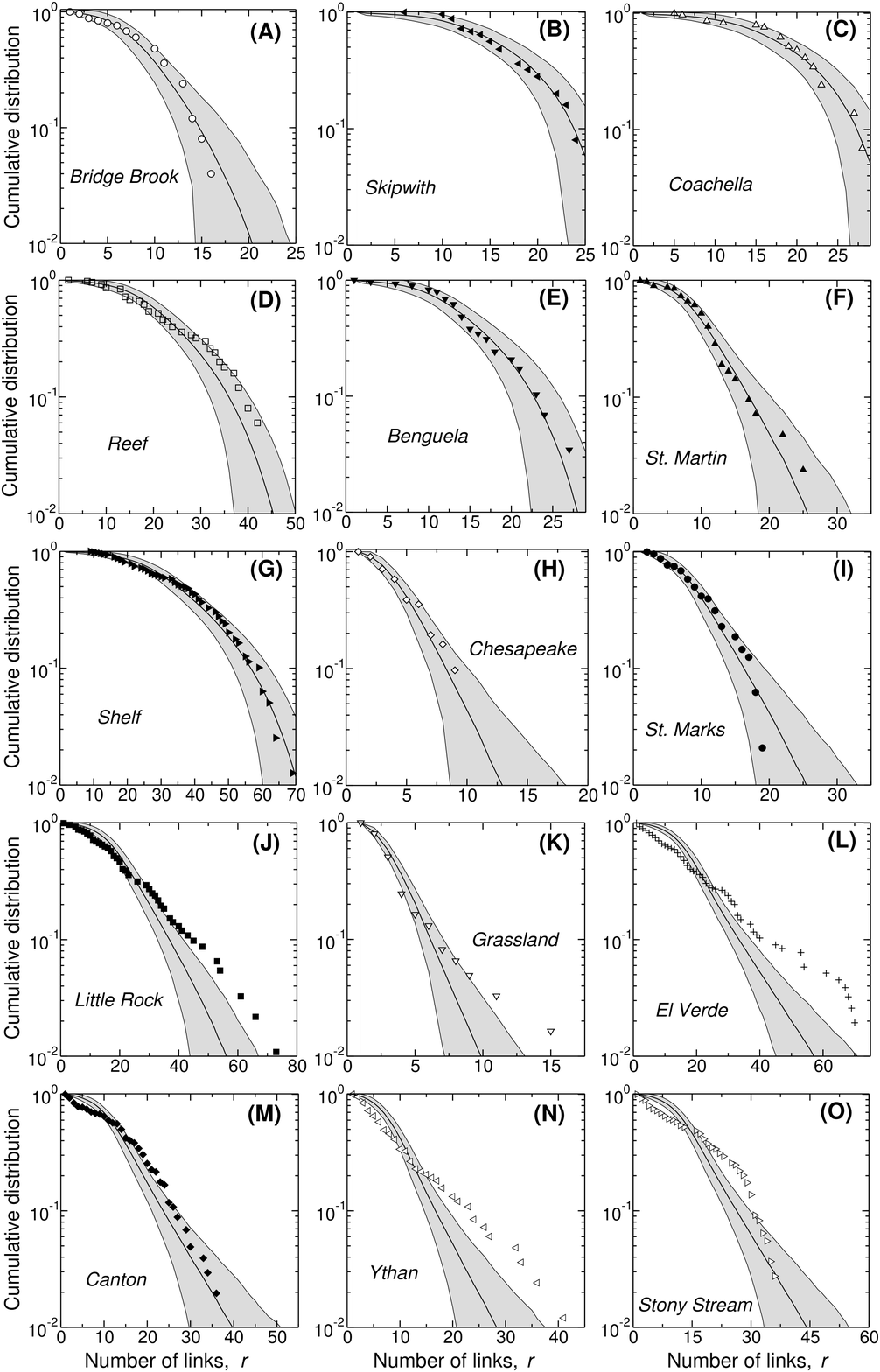}}
 \vspace*{-0.3cm}
 \caption{ Cumulative distribution $P_{\rm link}$ of the number $r$ of
 trophic links for the fifteen food webs studied (see
 Fig.~\protect\ref{preypanel}). The solid black line represents the
 average value from 1000 simulations of the niche model and the grey
 region represents two standard deviations above and below the model's
 predictions.\vspace{0cm}}
 \label{degreepanel}
\end{figure}

Next, we analyze the empirical data for fifteen food webs with 25 to
155 trophic species. These webs have linkage densities $1.6 < z <
17.7$, and connectances in the interval $0.026$--$0.315$ \citep
{martinez00,dunne02}. We first investigate the distributions of number
of prey and number of predators. In
Figs.{~\ref{preypanel}--\ref{degreepanel}} we compare the cumulative
distributions of the number of prey, number of predators, and number
of trophic links for species in the fifteen food web investigated with
the numerical predictions of the model for their specific values of
$S$ and $z$.


\begin{figure}[t]
 \vspace*{0.cm}
 \centerline{\includegraphics*[width=0.6\columnwidth]{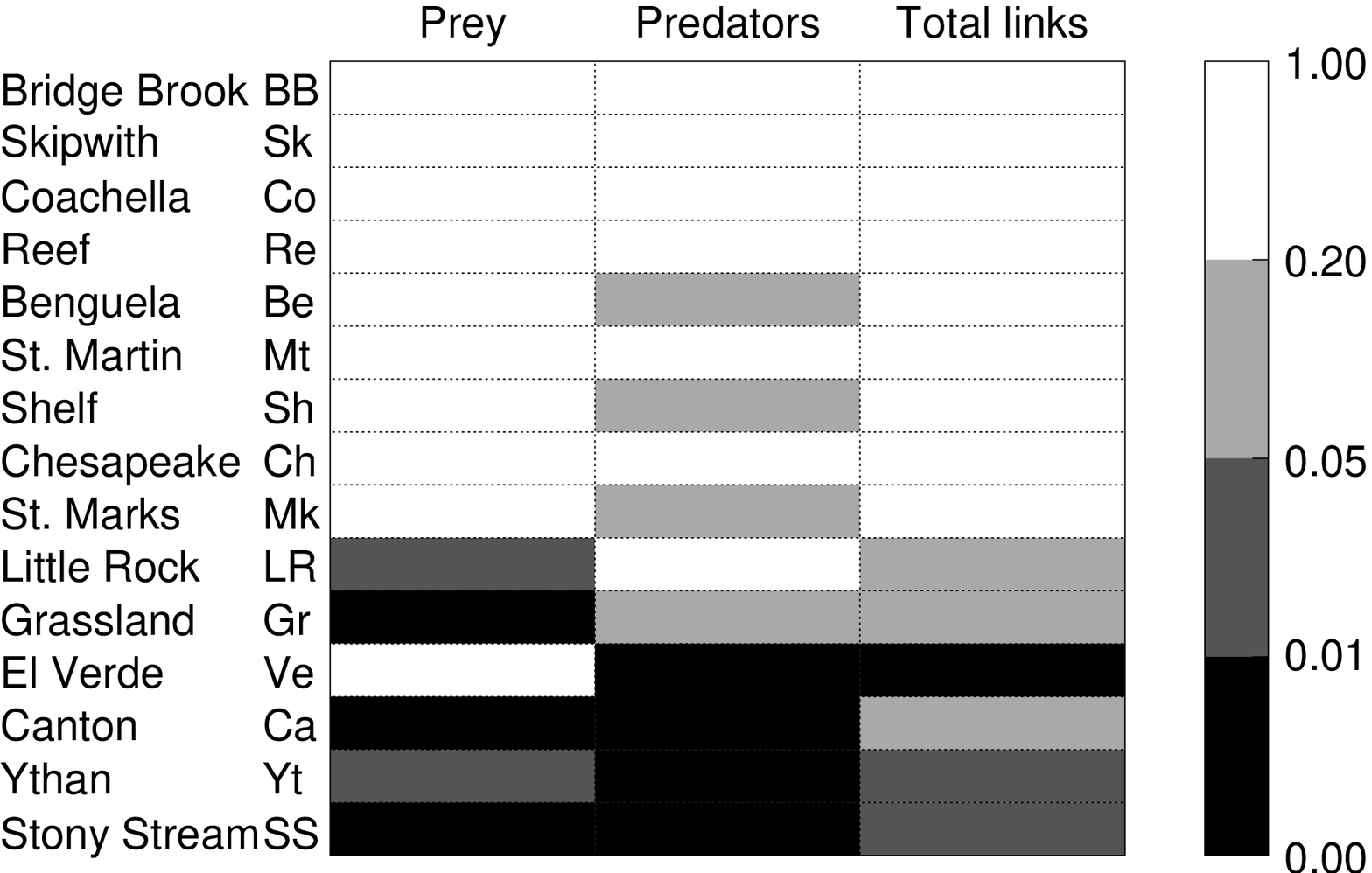}}
 \vspace*{-0.3cm}
 \caption{Comparison of the distributions of prey, predators, and
 total number of links of the fifteen food webs to the respective
 distributions obtained from 1000 webs generated by the niche
 model. We use the KS test for the comparison. Column one is the
 result for distribution of prey, column two the result for
 distribution of predators, and column three the result for
 distribution of total number of links. We regard ${\overline{P}_{\rm
 KS}} \le 0.01$---shown in black---as strong evidence for the
 rejection of the null hypothesis.\vspace{0cm}}
 \label{kstestvsmodel}
\end{figure}

To quantify the agreement between the empirical data and the model, we
apply the Kolmogorov-Smirnov (KS) test to the empirical
distributions---prey, predator, and number of trophic links---and 1000
realizations of the model. The average output of the KS test is
plotted in Fig.~\ref{kstestvsmodel}. We regard ${\overline{P}_{\rm
KS}} \le 0.01$---shown in black in Fig.~\ref{kstestvsmodel}---as
strong evidence for the rejection of the null hypothesis. Our results
suggest that eleven of the fifteen food webs studied are well
approximated by the niche model: Bridge Brook, Skipwith, Coachella,
Caribbean Reef, Benguela, St. Martin, Shelf, Chesapeake, St. Marks,
Little Rock, and Grassland. The remaining four---El Verde, Canton,
Ythan, and Stony Stream---exhibit rather different behavior, which is
visually apparent in Figs.~{\ref{preypanel}}--{\ref{degreepanel}} and
confirmed by the results in Fig.~\ref{kstestvsmodel}.

Equations~(\ref{cumprey})--(\ref{predcum2}) and the results of
Figs.~{\ref{preypanel}}--{\ref{kstestvsmodel}} suggest the possibility
that, to first approximation, $P_{\rm prey}$ and $P_{\rm pred}$ obey
universal functional forms that depend only on $z$. In order to
further investigate whether these patterns of similarity are true, as
well as if these distributions are universal, we perform the
Kolmogorov-Smirnov test between all pairs of food webs. To do this we
take advantage of the scaling determined earlier and use the scaled
variables as the basis for comparison. Equation (\ref{cumprey2})
predicts that $P_{{\rm prey}}(>k)$ depends only on $k/2z$. The scaling
of $P_{{\rm pred}}(>m)$ is not as straightforward. As discussed before,
``true'' scaling holds only for $m/2z < 1/2$, while for larger values
of $m/2z$ there is a Gaussian decay of the probability function with
an explicit dependence on $z$. However, the decay for $m>z$ is quite
fast and, to first approximation, not very relevant.

\begin{figure}[t]
 \vspace*{0.cm}
 \centerline{\includegraphics[width=0.85\columnwidth]{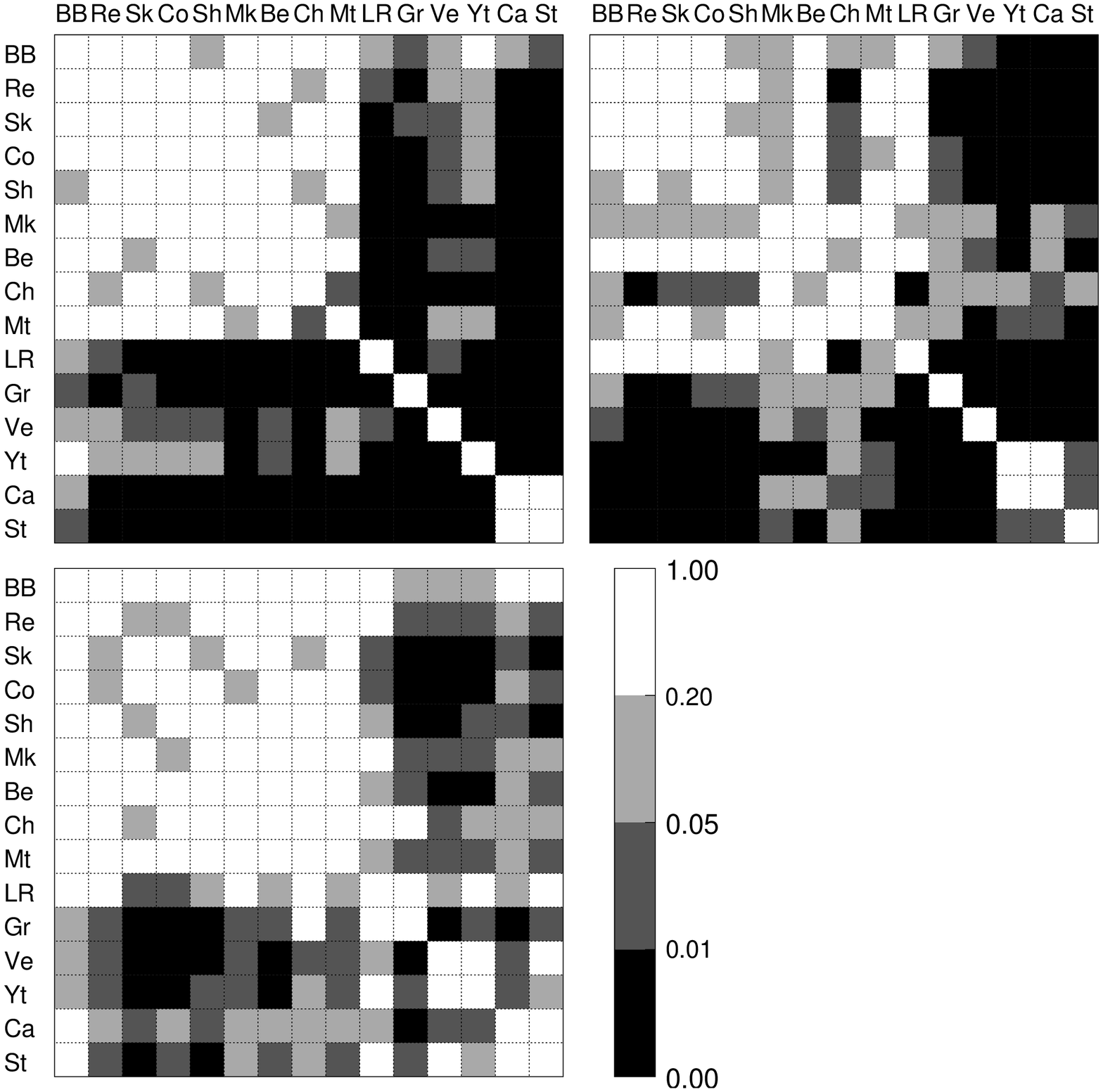}}
 \vspace*{-0.3cm}
 \caption{Comparison of the distributions of prey, predators, and
 total number of links of the fifteen empirical food webs. We use the
 KS test for the comparison. The resulting matrices may be interpreted
 as similarity matrices with values $0 \le P_{\rm KS} \le 1$, the KS
 probability. We show the matrix for distribution of prey $k$ in the
 top left, the matrix for distribution of predators $m$ in the top
 right, and the matrix for distribution of total number of links $r$
 in the bottom left. We regard ${\overline{P}_{\rm KS}} \le
 0.01$---shown in black---as strong evidence for the rejection of the
 null hypothesis.\vspace{0cm}}
 \label{correlationmatrices}
\end{figure}

We show the results of these tests in Fig.~\ref{correlationmatrices}.
We use the same rejection criterion as before. The results of
Fig.~\ref{correlationmatrices} also support the hypothesis that there
are eleven food webs that conform to the theoretical predictions.

For this reason, from this point on we will focus our attention upon
these eleven food webs---Bridge Brook, Skipwith, Coachella, Reef,
Benguela, St. Martin, Shelf, Chesapeake, St. Marks, Little Rock, and
Grassland.

\begin{figure}[t]
 \vspace*{0.cm}
 \centerline{\includegraphics*[width=\columnwidth]{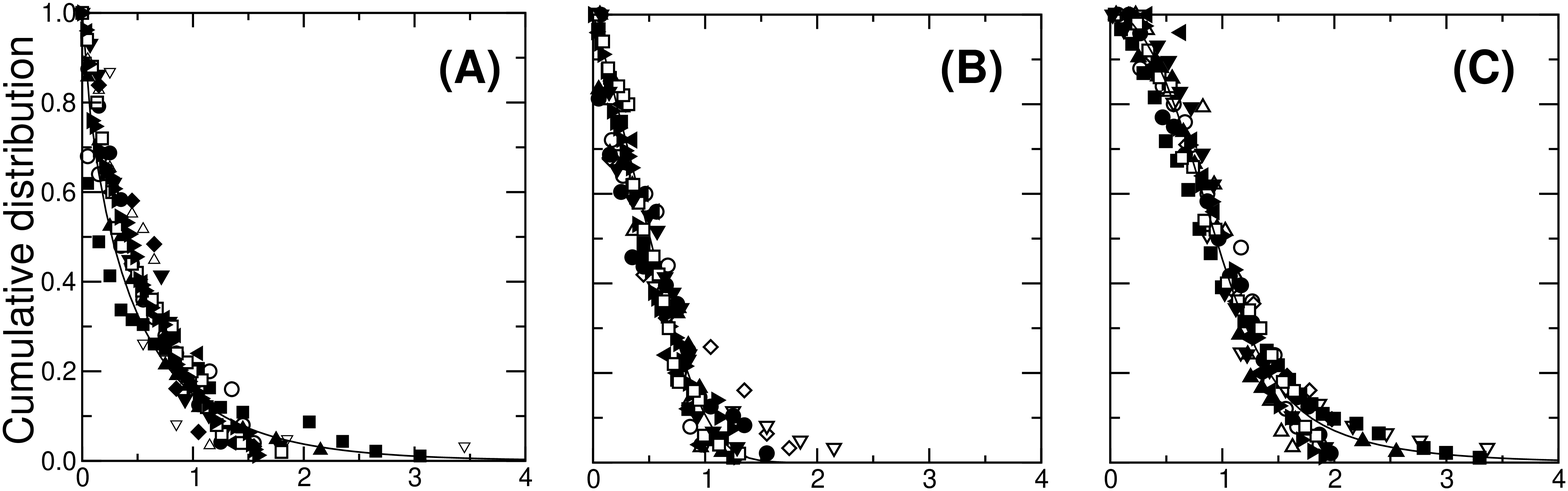}}
 \vspace*{-0.5cm}
 \centerline{\includegraphics*[width=\columnwidth]{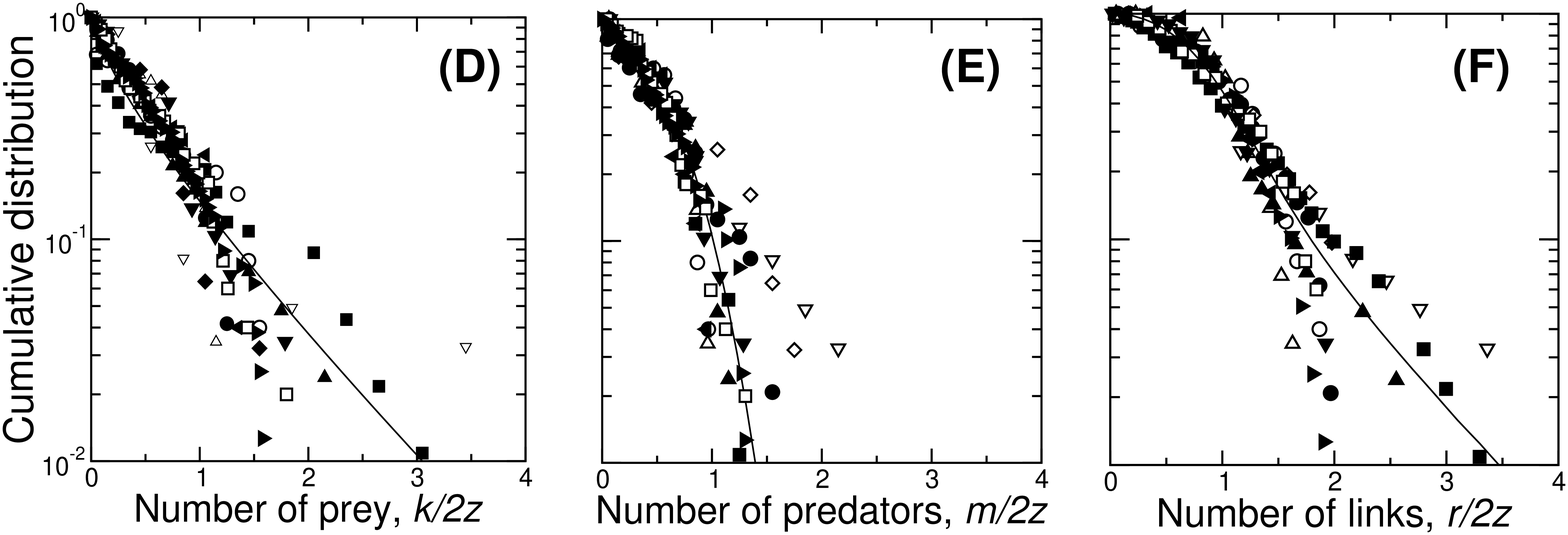}}
 \vspace*{-0.3cm}
 \caption{Visual test of the ``scaling hypothesis'' that the
 distributions of number of prey, predators, and trophic links have
 the same functional form for different food webs. (A) Cumulative
 distribution $P_{\rm prey}$ of the scaled number of prey $k/2z$ for
 the eleven food webs. The solid line is the prediction of
 Eq.~(\ref{cumprey}). The data ``collapse'' onto a single curve that
 agrees well with the analytical results. (B) Cumulative distribution
 $P_{\rm pred}$ of the scaled number of predators $m/2z$ for the
 eleven webs. The solid line is the analytical prediction of
 Eq.~(\protect\ref{predcum}) for the average value of $z$ in the
 empirical data, $z=8.44$ (C) Cumulative distribution $P_{\rm links}$
 of the scaled number of predators $r/2z$ for the eleven webs. The
 solid line is the analytical prediction. Semi-logarithmic plot of the
 scaled distributions of (D) number of prey, (E) number of predators,
 and (F) total number of links. The symbols are those introduced in
 Fig.~\protect\ref{preypanel}.\vspace{0cm}}
 \label{allunpooled}
\end{figure}

We plot in Figs.~\ref{allunpooled}(A) and (B) the cumulative
distributions $P_{{\rm prey}}(>k)$ versus the scaled variable $k/2z$
for the eleven similar food webs and find that the data collapse onto
a single curve, again supporting the possibility that $P_{{\rm prey}}$
obeys a universal functional form. Similarly we plot $P_{\rm pred}(>m)$
versus the scaled variable $m/2z$ for the eleven similar food webs in
Figs.~\ref{allunpooled}(B) and (C), again finding a collapse of the
data onto a single curve for $m/2z<0.5$.

\begin{figure}[t]
 \vspace*{0.cm}
 \centerline{\includegraphics*[width=\columnwidth]{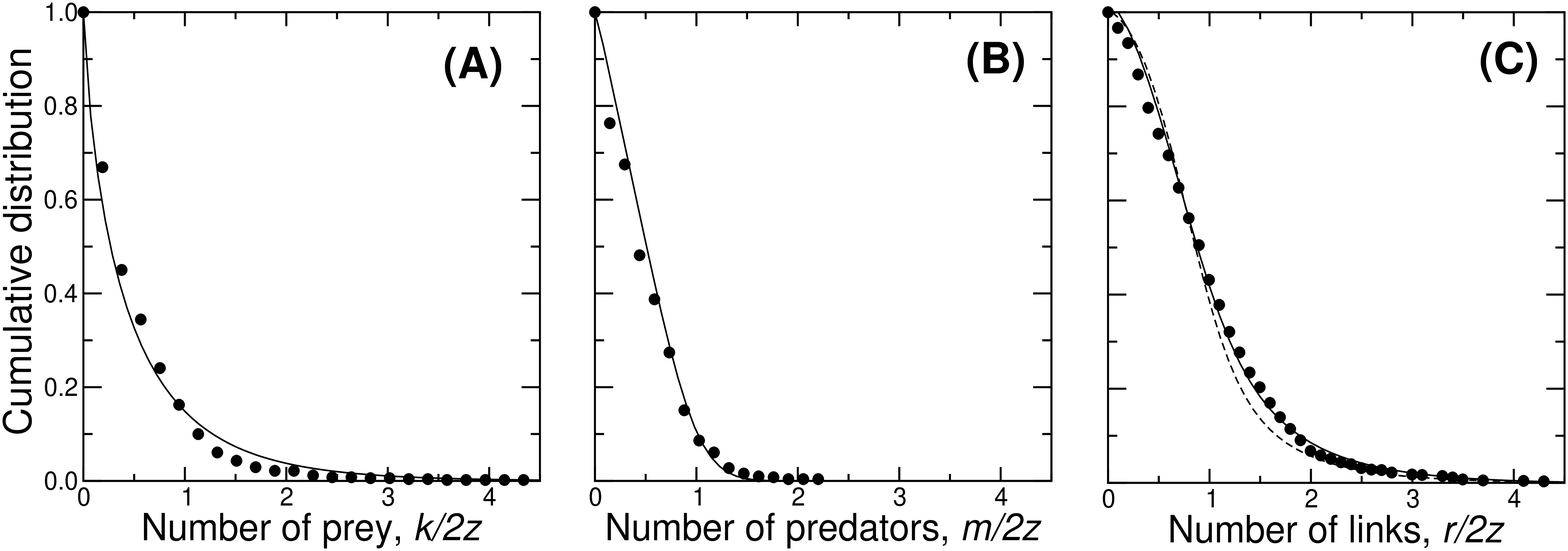}}
 \caption{Cumulative distributions (A) $P_{\rm prey}$ of the scaled
 number of prey, $k/2z$, and (B) $P_{{\rm pred}}$ of the scaled number
 of predators, $m/2z$, for the eleven pooled webs. The solid lines
 are, respectively, the analytical predictions Eq.~(\ref{cumprey2})
 and (\ref{predcum}), the latter with the average value $z=8.44$. (C)
 Cumulative distribution of the number of trophic interactions
 per species $r=k+m$ for the eleven pooled webs. The solid line is
 obtained by numerically convolving the distributions
 Eq.~(\ref{cumprey2}) and (\ref{predcum}) while the dashed line is
 obtained by numerical simulations of the niche model for $S=511$ and
 $z=8.44$, the parameter values of the pooled distributions. The tail
 of the distribution decays exponentially, indicating that food webs
 do {\it not\/} have a scale-free structure.}
 \label{pooledfigures}
\end{figure}

Figure~\ref{allunpooled} again supports the strong new hypothesis that
the distributions of number of prey and number of predators follow
{\it universal functional forms}. To improve statistics and better
investigate the specific functional form of these distributions, one
may pool the scaled variables, $k/2z$ and $m/2z$, from the different
webs into single distributions, $P_{{\rm prey}}$ and $P_{{\rm pred}}$,
respectively. Figures~\ref{pooledfigures}(A) and (B) show the
cumulative distributions of scaled number of prey and scaled number of
predators for the pooled webs. Note that the distributions are well
approximated by Eqs.~(\ref{cumprey2}) and (\ref{predcum}) {\it even
though there are no free parameters to fit in the analytical
curves}. These results are analogous to the finding of scaling and
universality in physical, chemical, and social systems
\citep{stanley71,stanley99,bunde94}.

In Fig.~\ref{pooledfigures} we plot the probability densities for the
distribution of number of prey and number of predators for the pooled
webs. It is visually apparent that both distributions are
different. This is confirmed by the Kolmogorov-Smirnov test which
rejects the null hypothesis at the 0.2\% level. The distribution of
number of prey decays exponentially while the distribution of number
of predators is essentially a step function.

One can perform a similar analysis for the distribution $p_{{\rm
link}}$ of the number of trophic links $r$. As for number of prey or
number of predators, the data from the different webs, upon the
scaling $r/2z$, collapse onto a single curve, further supporting the
hypothesis that scaling holds for food web structure. To better
determine the specific functional form of $p_{\rm link}(r)$, in
Fig.~\ref{pooledfigures}(D) we pool the scaled variables, $r/2z$, from
the eleven food webs into a single distribution. We find that $p_{\rm
link}(r)$ has an exponential decay for $r/2z \gg 1$, in agreement with
our theoretical calculations. Therefore, there is a characteristic
scale for the linkage density, i.e. food webs do {\it not\/} have a
scale-free structure, in contrast to erroneous reports in recent
studies of food-web fragility \citep{sole01,montoya02}.

\begin{figure}[t!]
 \vspace*{0.cm}
 \centerline{\includegraphics*[width=0.5\columnwidth]{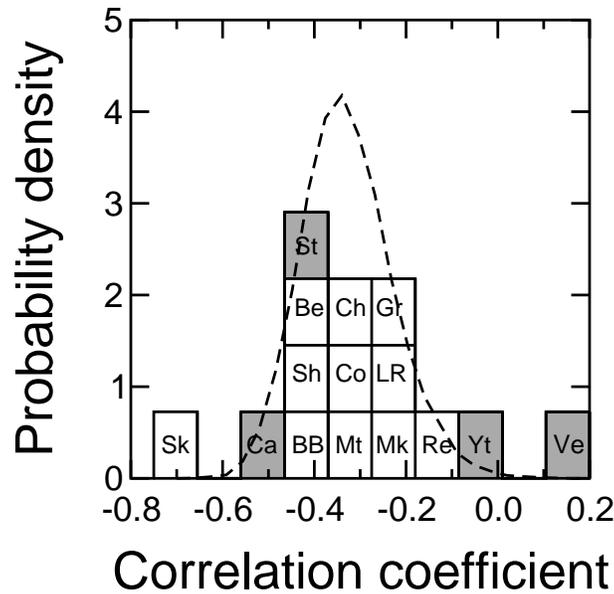}}
 \vspace*{-0.3cm}
 \caption{Correlation coefficient between number of prey and number of
 predators for the fifteen food webs investigated. Probability density
 plot for the correlation coefficients obtained empirically compared
 to the probability density of $r_{\rm cor}$ from 5000 realizations of
 the niche model at the average properties of the empirical webs
 studied, $S=64$ and $z=8.06$. Note the good agreement between data
 and model predictions. The white boxes represent the eleven food webs
 which are well described by our analytical expressions and the niche
 model while the grey boxes represent the four food webs which are
 not.
}
 \label{correlationcoeff}
\end{figure}

The analytical expression for the distribution of the total number of
links, Eq.~(\ref{convol}), was derived assuming that the number of
prey and number of predators are uncorrelated. In
Fig.~\ref{correlationcoeff} we compare the empirical $r_{\rm cor}$
with data generated from simulations of the niche model. It can be
seen that, just as in the niche model, food webs show, on average,
slightly negative correlation between number of prey and number of
predators. It is also important to note that for both the empirical
data and the niche model, the is a wide range of possible correlation
coefficients which may occur.

\subsection{Assortativity}

\begin{figure}[t!]
 \vspace*{0.cm}
 \centerline{\includegraphics*[width=0.5\columnwidth]{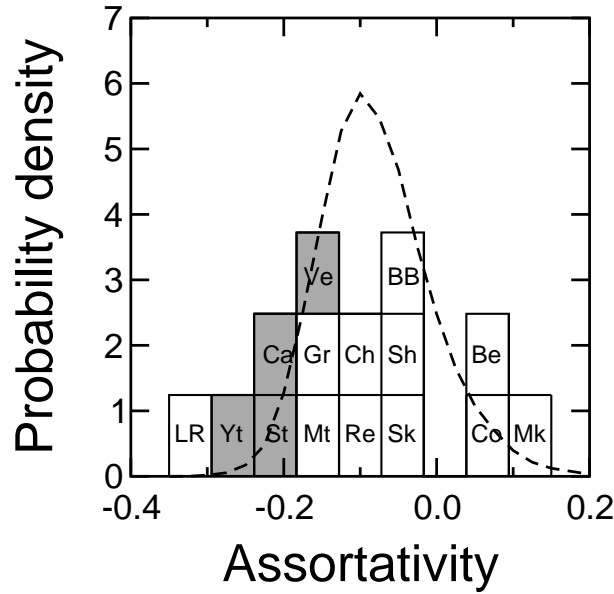}}
 \vspace*{-0.3cm}
 \caption{Assortativity of the fifteen food webs
 investigated. Probability density plot for the assortativities
 obtained empirically compared to the probability density of
 assortativity from 5000 realizations of the niche model at the
 average properties of the empirical webs studied, namely $S=64$ and
 $z=8.06$. The food webs are, on average, slightly disassortative;
 however, they cover a wide range of values from moderately
 disassortative to slightly assortative and all of these behaviors are
 captured in the predictions of the model. Assortativities were
 calculated as outlined in \protect\citep{newman02}. As in
 Fig.~\ref{correlationcoeff}, the white boxes represent the eleven
 food webs which are well described by our analytical expressions and
 the niche model while the grey boxes represent the four food webs
 which are not.\vspace{0cm}}
 \label{assortativity}
\end{figure}

It has been reported that technological and biological networks are
most commonly disassortive, in contrast to social networks which are
most commonly assortative in nature \citep{newman02}. We analyze the
assortativities of the fifteen empirical food webs investigated and
compare this data to numerical simulations of the model. We confirm
that food webs are mostly disassortative. However, the assortativities
cover a wide range of values from moderately disassortative to
slightly assortative. The numerical results from simulations of the
niche model also show a similarly wide range of possible values, as
shown in Fig.~\ref{assortativity}.

\subsection{Cannibalism}

\begin{figure}[t!]
 \vspace*{0.cm}
 \centerline{\includegraphics*[width=0.85\columnwidth]{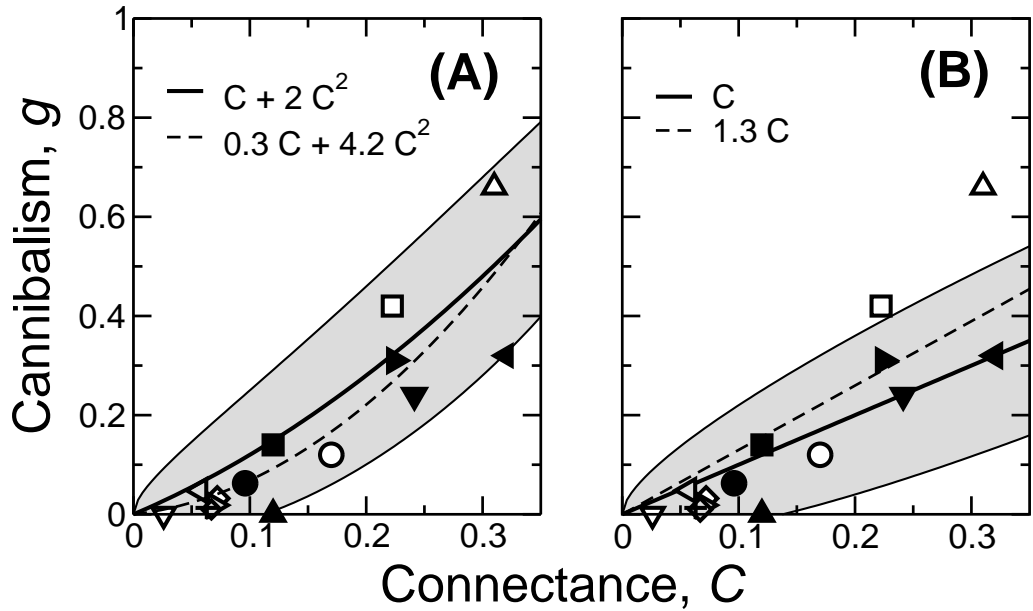}}
 \vspace*{-0.3cm}

 \caption{Fraction of cannibal species as a function of the
 connectance $C$ for empirical webs relative to (A) the niche model
 and (B) the random model. Thick solid lines correspond to model
 predictions Eqs. (\ref{can2}) and (\ref{canrandom}), respectively.
 The shaded area is within two standard deviations from the model
 average. One observes that the empirical results are in good
 agreement with the niche model. The dashed lines represent the best
 fit quadratic (A) and linear (B) expressions for the empirical data.
 The symbols are those introduced in Fig.~\protect\ref{preypanel}
 except that the four poorly approximated food webs are filled in
 grey.\vspace{0cm}}
 \label{cann_new}
\end{figure}

We determine the fraction of cannibal species as a function of the
connectance $C$ for the empirical webs considered, and compare it with
the predictions of the niche and the random model analyzed in
Appendix \ref{cannappendix}. We find that the analytical expression for
cannibalism supplied by the niche model captures the behavior of all
fifteen food webs we study. In contrast, the predictions of the random
model, shown in Fig.~\ref{cann_new}(B), cannot capture the behavior of
Coachella or the Caribbean Reef webs.

It is observed that the empirical values are generally smaller than
the predictions. Indeed, some food webs have no cannibals at all. Such
low ratios of cannibalism is seen by some authors as unrealistic, and
possibly due to the large effort needed to build a comprehensive food
web \citep{polis91}.
%
%
%


\subsection{Network theory mesaures}

Next, we investigate if the scaling hypothesis suggested by the
analysis of distribution of trophic links also applies to other
quantities characterizing food web structure. We consider two
quantities with ecologic implications: (i) the average trophic
distance $d$ between species \citep{watts98}, which is defined as the
typical number of species needed to trophically connect two given
species, and (ii) the clustering coefficient $\mathcal C$
\citep{watts98} which quantifies the fraction of species triplets that
form fully-connected triangles.

\begin{figure}[t]
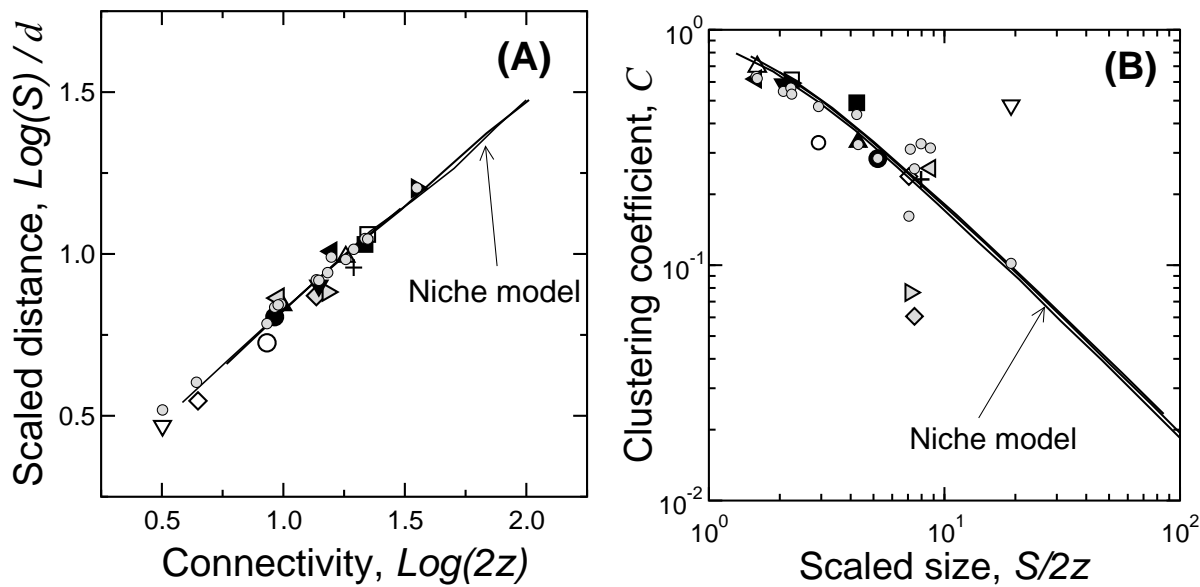

 \vspace*{0.cm}
 \centerline{\includegraphics*[width=0.48\columnwidth]{l_z_all}\quad\includegraphics*[width=0.48\columnwidth]{clustering}}
 \vspace*{-0.3cm}
 \caption{ (A) Scaled average trophic distance $d$ between species
 versus linkage density $z$. We compare the data with the numerical
 simulations of the niche model \protect\citep{martinez00} for web
 sizes $S=100, 500, 1000$ (thin solid lines). We find a logarithmic
 increase of the average distance with web size $S$ for the empirical
 food webs, in good agreement with the model predictions.
 (B) Double-logarithmic plot of the clustering coefficient $\mathcal
 C$ versus the scaled web size $S/2z$. We compare the data with
 numerical results for the niche model \protect\citep{martinez00} for
 three values of the linkage density in the empirically-relevant range
 ($z = 2.5, 5, 10$). We find that the clustering coefficient of the
 food webs is inversely proportional to the web size $S$, in good
 agreement with the model predictions and with the asymptotic behavior
 predicted for a random graph \protect\citep{watts98}. For both (A)
 and (B) the grey circles represent the average values calculated from
 1000 randomizations of the links of the empirical food webs keeping
 the same distributions of number of prey and number of
 predators. Note that the behavior of these randomized webs is still
 captured quite well by the niche model implying that the underlying
 distributions themselves are responsible for this behavior. The
 symbols are those introduced in Fig.~\ref{preypanel} except that the
 four poorly approximated food webs are filled in grey.}
 \label{distanceandclustering}
\end{figure}

In Fig.~\ref{distanceandclustering}(A) we compare our numerical
results for the average trophic distance $d$ for the niche model
\citep{martinez00} with the values calculated for the food webs
analyzed. We observe that there is agreement between the niche model
and the empirical data. Remarkably, the behavior predicted by the
model also holds for randomization of the empirical data, where the
randomization of links between species is such that the distributions
of number of predators and number of prey remain the same. We also
find that $d$ increases with web size as $\log S$ both for the model
and for the data.
%

The results of Fig.~\ref{distanceandclustering}(A) also support the
scaling hypothesis and suggest that the average distance in a food web
may also follow a unique functional form for different food webs.
This feature is predicted remarkably well by the niche model, even for
the four webs for which the model fails to describe other topological
properties.

Figure~\ref{distanceandclustering}(B) shows our results for the
clustering coefficient $\mathcal C$ of the food webs studied and for
the niche model. We find that the data is well approximated by the
model predictions which show that $\mathcal C$ decreases to zero as
$1/S$ as web size $S$ increases. This result does not hold for
Grassland---which has a much higher value than predicted---nor for
Canton or Stony Stream---which feature much lower values than
predicted. As for the average distance $d$, the niche model accurately
predicts the values of the clustering coefficient even after the
randomization of the empirical food webs. Again, this suggests that
there are robust features inherent in the simple framework of the
niche model which allows it to describe well the most complete
empirical data available.

\section{Concluding remarks}

The major finding of this study is the uncovering of unifying
quantitative patterns characterizing the structure of food webs from
diverse environments. Specifically, we find that, for the majority of
the most complete empirical food webs, the distributions of the number
of prey, number of predators, and number of trophic links obey
universal scaling functions, where the scaling quantity is the linkage
density. Remarkably, these scaling functions are consistent with
analytical predictions we derived for the niche model. Therefore, our
results suggest that these distributions can be theoretically
predicted merely by knowing the food web's linkage density, a
parameter readily accessible empirically.

Our results are of interest for a number of reasons. 
First, the results of Fig.~\ref{distanceandclustering}---which also
support the scaling hypothesis---indicate that there is very little,
if any, compartmentalization in ecosystems \citep{Pimm80}, suggesting
the possibility that ecosystems are highly interconnected and that the
removal of any species may induce large disturbances.
Second, regularities such as these are interesting as descriptors of
trophic interactions inside communities because they may enable us to
make predictions in the absence of high-quality data, and provide
insight into how ecologic communities function and are assembled.
Third, the structure of food webs is different from that of many other
biological networks in two important aspects: the links are
uni-directional and the in- and out-degree distribution are different.
These two facts are a result of the {\it directed\/} character of the
trophic interactions and of the asymmetry it creates. Interestingly,
the niche model captures this asymmetry in its rules, which may
explain its success in explaining the empirical results.
Fourth, food webs do not have a scale free distribution of number of
links (total, prey, or predators). This may be viewed as surprising
since one could expect most species to try to prey on the most
abundant species in the ecosystem (an ``abundant-get-eaten'' kind of
mechanism). Such a preferential attachment would lead to a scale-free
distribution of links; instead, we find a single-scale distribution,
suggesting that species specialize and prey on a small set of other
species.

We do not intend to provide here a detailed study of why four of the
fifteen food webs we consider do not accomodate to the patterns we
report. We will, nonetheless, offer a few remarks on this matter. The
four webs that do not conform to our patterns are Ythan Estuary,
Canton Creek, Stony Stream, and El Verde Rainforest. Let us first
consider the Ythan Estuary web. It has been noted already that this
web displays an over-representation of top bird species
\citep{martinez00} which could account for the differences. For Canton
Creek and Stony Stream two aspects distinguish them from the other
food webs. First, they are quantitatively quite similar. This fact is
illustrated by noting that the Kolmogorov-Smirnov test provides
${P_{\rm KS}}$ of $0.565, 0.045,$ and $0.794$ under direct comparison
of their respective distributions of number of prey, number of
predators, and total number of links. Upon further investigation this
easily explained as the original authors stated plainly that it was
their intention to compile food webs from habitats that were as
similar as possible to each other \citep{townsend98}. Second, the
Canton Creek and Stony Stream webs are also time specific
\citep{townsend98}--meaning their data was collected solely on one
occasion---as opposed to cumulative---which are based upon data
collected on multiple occasion until reaching some state of
``completeness.'' This time specific nature not only distinguishes
them from all of the other food webs which we have studied, but also
implies that they contain rather different information and are not
directly comparable to cumulative food webs in our framework. We
lastly address the El Verde Rainforest food web by pointing out that
over one third of the links were not observered in the field but
rather are based upon iteractions involving closely related species in
the forest or observations or published accounts of the interaction
outside of the forest \citep{waide96}.

To conclude, we want to stress that our findings are remarkable for
two main reasons: (i) they hold for eleven out of fifteen of the most
complete food webs studied, in contrast to previously reported
patterns, and (ii) they support the conclusion that fundamental
concepts of modern statistical physics such as scaling and
universality---which were developed for the study of inanimate
systems---may be succesfully applied in the study of food webs---which
comprise animate beings
\citep{camacho02a,camacho02b,caldarelli03}. Indeed, our results are
consistent with the underlying hypothesis of scaling theory, i.e.,
food webs display ``universal'' patterns in the way trophic relations
are established despite apparently ``fundamental'' differences in
factors such as the environment (e.g. marine versus terrestrial),
ecosystem assembly, and past history.

\vspace{2cm} \centerline{Acknowledgments} We thank J. Bafaluy,
M. C. Barbosa, M. Barth\'el\'emy, J. Faraudo, G. Franzese, S. Mossa,
R. V. Sol\'e, H. E. Stanley, and especially N. D. Martinez and
R. J. Williams for stimulating discussions and helpful suggestions. We
also thank J. E. Cohen, N. D. Martinez, R. J. Williams, and J. Dunne
for making their electronic databases of food webs available to us. JC
thanks the Spanish CICYT (BFM2003-06033) for support. RG thanks the
Fulbright Program and the Spanish Ministry of Education, Culture \&
Sports. DBS thanks the NU ChBE Murphy Fellowship and the NSF-IGERT
program "Dynamics of Complex Systems in Science and Engineering"
(DGE-9987577).





\appendix

\section{Variables}

\begin{tabular}{cl}
\hline\vspace{-.7cm}\vspace{-.1cm}\\
Model & Variable \vspace{-.1cm}\\
variable & definition \vspace{-.1cm}\\
\hline\hline\vspace{-.7cm}\vspace{-.1cm}\\
$S$ & Number of trophic species \vspace{-.1cm}\\
$L$ & Total number of trophic links \vspace{-.1cm}\\
$C$ & Directed connectance, $C=L/S^2$\vspace{-.1cm}\\
$z$ & Linkage density, $z=L/S$\vspace{-.1cm}\\
$n$ & Niche number in niche model, $n \in [0,1]$\vspace{-.1cm}\\
$a$ & Range of predation in niche model\vspace{-.1cm}\\
$x$ & Beta-distributed random variable\vspace{-.1cm}\\
$b$ & Characteristic parameter for beta-distribution of $x$\\
\hline
\end{tabular}

\begin{tabular}{cl}
\hline\vspace{-.7cm}\vspace{-.1cm}\\
Ecological & Variable \vspace{-.1cm}\\
variable & definition \vspace{-.1cm}\\
\hline\hline\vspace{-.7cm}\vspace{-.1cm}\\
$k$ & Number of prey\vspace{-.1cm}\\
$\tilde{k}$ & Scaled number of prey, $\tilde{k} \equiv k/2z$\vspace{-.1cm}\\
$m$ & Number of predators\vspace{-.1cm}\\
$\tilde{m}$ & Scaled number of predators, $\tilde{m} \equiv m/2z$\vspace{-.1cm}\\
$r$ & Total number of trophic links, $r \equiv k+m$\vspace{-.1cm}\\
$\tilde{r}$ & Scaled total number of trophic links, $\tilde{r} \equiv r/2z$\vspace{-.1cm}\\
$T$ & Fraction of top species \vspace{-.1cm}\\
$B$ & Fraction of basal species \vspace{-.1cm}\\
$I$ & Fraction of intermediate species \vspace{-.1cm}\\
$g$ & Fraction of cannibals \vspace{-.1cm}\\
$\sigma_{V}$ & Standard deviation of vulnerability\vspace{-.1cm}\\
$\sigma_{G}$ & Standard deviation of generality\vspace{-.1cm}\\
$r_{\rm cor}$ & Correlation coefficient between species number\vspace{-.1cm}\\
 & of prey and number of predators\vspace{-.1cm}\\
$A$ & Assortativity\\ 
\hline
\end{tabular}

\section{Fraction of cannibal species for a random linking model}
\label{cannappendix}

In this Appendix we calculate the fraction
of cannibal species for a random model, i.e. a model where the
links between species are placed at random. For a system with $S$
species and $z$ average number of links per species, the probability
that a species feeds on itself will be
\begin{equation}
g_{\rm rand}=\frac{z}{S}=C.
\label{canrandom} 
\end{equation}
The standard deviation of $g_{\rm rand}$ is obtained from
Eq.~(\ref{deltag}) by using $g=g_{\rm rand}$, yielding
\begin{equation}
\sigma_{g_{\rm rand}}=\sqrt{\frac{g_{\rm rand}(1-g_{\rm rand})}{S}}\,.
\label{deltagrand}
\end{equation}
In the limit of small $C$, the predictions of both models are
similar. Note that this occurs regardless of the particular
form of $p_{x}(x)$. In fact, $C \rightarrow 0$ is equivalent to
$\overline{x} \rightarrow 0$, so that $p_{x}(x)$ takes non-vanishing
values only for very small $x$. One can then neglect $x/2$ versus $1$
in the denominator of Eq.~(\ref{can0}) and get
\begin{equation}
g=\int_{0}^{1} p_{x}(x) \frac{x/2}{1-x/2} dx \simeq
\frac{\overline{x}}{2}=C=g_{\rm rand} \,.
\end{equation} 
Therefore, only the second term in the right hand side of expression
Eq.~(\ref{can2}) depends on the specific form of $p_{x}(x)$.

\section{Calculation of the moments of the distributions of number of prey and number of predators}

In this Appendix we calculate averages of the type
$\overline{k^{i}m^{j}}$, with $i$ and $j$ integers,
in the limit of large web sizes and small connectances. These
averages are used in the calculation of the assortativity, in
Section \ref{secassor}.

In Section \ref{seccor}, we obtained Eq. (\ref{pkm}) for the join
probability $p(k,m)$, namely
\begin{equation} p(k,m)=\int dn \, p(k,m,n) = \int dn \, p(k|n) p(m|n) \,.
\label{apkm}
\end{equation}
With this, the average we want to calculate writes
\begin{equation}
 \overline{k^{i}m^{j}}= \int_{0}^{1}dn \left(\sum_{k=0}^{S}
 k^{i} p(k|n)\right) \left(\sum_{m=0}^{S} m^{j}
 p(m|n)\right) \equiv \int_{0}^{1}dn \, \overline{k_{n}^{i}} \,
 \overline{m_{n}^{j}} \,.
\label{kmab}
\end{equation}

The probability $p(k|n)$ of a species with niche value $n$ to have $k$
prey can be calculated as follows. According to the rules of the niche
model, in order to have $k$ prey, the parameter $x$ must take the
value $x=k/Sn$. Then, by performing a change of variables, one obtains
\begin{equation}
p(k|n)=\frac{1}{n S}p_{x}(\frac{k}{n S}) \nonumber \,,
\end{equation}
where $p_{x}$ is given by Eq. (\ref{px}). In the limit of small
$C$, $p_{x}$ can be approximated by
\begin{equation}
p_{x}(x) \simeq b \exp(-bx) \simeq 2C \exp(-x/2C).
\end{equation}
Then, $p(k|n)$ is given by
\begin{equation}
p(k|n)= \frac{1}{2nz} \exp(-\frac{k}{2nz}) \,.
\label{pkn}
\end{equation}
The average values $\overline{k_{n}^{i}}$ can be evaluated, in
the limit of small $C$, as
\begin{equation}
\overline{k_{n}^{i}}=\int_{0}^{\infty} k^{i} \frac{1}{2nz} \exp(-\frac{k}{2nz}) = 
\left(2nz \right)^{i} {i}!
\label{ka}
\end{equation}
In order to evaluate $\overline{m_{n}^{j}}$ we need $p(m|n)$. In
the limit $S\gg 1$ and $C\ll 1$, the number of potencial predators of
a species of niche value $n$ (that is, the number of species with a
niche number larger than $n$) is $S(1-n)$. As seen in Section
\ref{secnpred}, each of them has a probability $\overline{x}$ to prey
on the species. Thus, the probability that a species with niche
parameter $n$ has $m$ predators is simply the binomial distribution,
\begin{equation}
p(m|n)={{S(1-n)}\choose m} \overline{x}^{m}
(1-\overline{x})^{S(1-n)-m},
\end{equation}
that in our limit yields the Poisson distribution, namely
\begin{equation}
p(m|n)=\frac{\lambda^{m}\exp{(-\lambda)}}{\lambda!},
\label{pmn}
\end{equation}
with 
\begin{equation}
\lambda=S(1-n)\overline{x}=2z(1-n) \,.
\label{lambda}
\end{equation}

One can now obtain the moments
\begin{equation}
\overline{m_{n}}= \lambda
\label{ms1}
\end{equation}
\begin{equation}
\overline{m_{n}^{2}}= \lambda^{2}+\lambda
\end{equation}
\begin{equation}
\overline{m_{n}^{3}}= \lambda^{3}+3 \lambda^{2}+\lambda \,.
\label{ms3}
\end{equation}

Finally, using Eq. (\ref{ka}), (\ref{lambda}) and
(\ref{ms1})--(\ref{ms3}) into Eq. (\ref{apkm}), yields
\begin{equation}
\overline{k}= \overline{m}= z\,,
\end{equation}
\begin{equation}
\overline{k^{2}}= 8 z^{2}/3\,,
\end{equation}
\begin{equation}
\overline{k^{3}}= 12 z^{3}\,,
\end{equation}
\begin{equation}
\overline{m^{2}}= z+4z^{2}/3\,,
\end{equation}
\begin{equation}
\overline{m^{3}}= z+4z^{2}+2z^{3}\,,
\end{equation}
\begin{equation}
\overline{k \, m}= 2z^{2}/3\,,
\end{equation}
\begin{equation}
\overline{k \, m^{2}}= 2z^{2}(1+z)/3\,,
\end{equation}
and
\begin{equation}
\overline{k^{2} m}= 4z^{2}/3.
\end{equation}

\end{document}